\documentclass[superscriptaddress,
twocolumn,
bibnotes,amsmath,amssymb,aps,pra,floatfix]{revtex4-2}

\usepackage[margin=1in]{geometry}
\usepackage[english]{babel}
\usepackage{xcolor}
\usepackage{ucs}
\usepackage[utf8]{inputenc}
\usepackage{amsmath}
\usepackage{amssymb}
\usepackage{amsthm}
\usepackage{graphicx}
\usepackage{bm}
\usepackage{bbm}
\usepackage{braket}
\usepackage{empheq}
\usepackage{hyperref}

\begin{document}

\title{Spin squeezing in mixed-dimensional anisotropic lattice models}
\date{\today}

\author{Mikhail Mamaev}
\email{mikhail.mamaev@colorado.edu}
\affiliation{JILA, NIST and Department of Physics, University of Colorado Boulder, Boulder, USA, 80309}
\author{Diego Barberena}
\affiliation{JILA, NIST and Department of Physics, University of Colorado Boulder, Boulder, USA, 80309}
\author{Ana Maria Rey}
\affiliation{JILA, NIST and Department of Physics, University of Colorado Boulder, Boulder, USA, 80309}

\begin{abstract}
{
We describe a theoretical scheme for generating scalable spin squeezing with nearest-neighbour interactions between spin-1/2 particles in a 3D lattice, which are naturally present in state-of-the-art 3D optical lattice clocks. We propose to use strong isotropic Heisenberg interactions within individual planes of the lattice, forcing the constituent spin-1/2s to behave as large collective spins. These large spins are then coupled with XXZ anisotropic interactions along a third direction of the lattice. This system can be realized via superexchange interactions in a 3D optical lattice subject to an external linear potential, such as gravity, and in the presence of spin-orbit coupling (SOC) to generate spin anisotropic interactions. We show there is a wide range of parameters in this setting where the spin squeezing improves with increasing system size even in the presence of holes.
}
\end{abstract}

\maketitle

\section{Introduction}

Ultracold atoms in optical lattices are a powerful tool for modern applications of quantum technology. One active pursuit is quantum-enhanced metrology in the form of spin squeezing~\cite{kitagawa1993squeezing,wineland1992squeezing}, which can push sensors beyond conventional limitations by exploiting many-body entanglement between the atoms. Many works have proposed theoretical methods to generate squeezing~\cite{ma2011squeezingReview,pezze2018squeezingReview}, and several experiments have succeeded at realizing proof-of-concept spin-squeezed states~\cite{takano2009squeezingExpBEC,appel2009squeezingExpBEC,gross2010squeezingExpBEC,riedel2010squeezingExpChip,schleier2010squeezingExpCavity,hamley2012squeezingExpBEC,bohnet2014squeezingExpCavity,hosten2016squeezingExpCavity,bohnet2016squeezingExpIons,braverman2019squeezingExpCavity,robinson2022squeezingExpCavity,eckner2023squeezingExpRydberg,bornet2023squeezingExpRydberg,hines2023squeezingExpRydberg,franke2023squeezingExpIons}. However, no experimental efforts have yet surpassed the absolute precision of state-of-the-art sensors using uncorrelated atoms.

One reason for this lack of progress is that scalable spin squeezing is easiest to generate in systems with long-range interactions, such as  dipolar~\cite{bilitewski2021squeezingThrDipole,gil2014squeezingThrRydberg,bornet2023squeezingExpRydberg,eckner2023squeezingExpRydberg,hines2023squeezingExpRydberg}, phonon~\cite{bohnet2016squeezingExpIons,franke2023squeezingExpIons}, and photon mediated ~\cite{groszkowski2020squeezingThrCavity,lewis2018squeezingThrCavity,bohnet2014squeezingExpCavity,hosten2016squeezingExpCavity,braverman2019squeezingExpCavity,robinson2022squeezingExpCavity} interactions, which are not naturally present in state-of-the-art 3D optical lattice clocks. The latter only have on-site collisional interactions and thus local connectivity, which is non-ideal for spin squeezing  generation~\cite{hazzard2014squeezingAndCorrelations,lee2013spinModelsSqueezing,foss2013ising}.

Recent studies have nevertheless suggested that in certain parameter regimes, short-range interacting systems can behave like collective long-ranged ones during transient quantum dynamics~\cite{he2019squeezing,foss2016squeezingShortRange,perlin2020spin,yanes2022squeezingThrSOC}. This behavior can remain valid long enough to generate significant squeezing that scales with system size, even for nearest-neighbour models such as superexchange interactions in 3D optical lattices. Unfortunately, for nearest-neighbour spin-1/2 models, the collective regime is parametrically restrictive and requires XXZ-type spin interactions with anisotropy $\Delta$ that satisfies $|\Delta - 1| \ll 1$, near the Heisenberg point $\Delta = 1$, preferably in the easy-plane phase $|\Delta| < 1$. This leads to slow timescales as the squeezing time is proportional to $1/|\Delta-1|$. Furthermore, holes in the initial state are unavoidable at finite entropy and create further challenges due to their fast propagation rates compared to the superexchange interactions, which can disrupt the spin dynamics.

In this work we propose to overcome these challenges by generating XXZ-type models which  behave as a large-spin chain. The idea is to generate  strong  isotropic Heisenberg interactions across 2D planes of an optical lattice. These interactions lock the constituent spin-1/2s in each plane into a large collective  spin~\cite{rey2008collective,cappellaro2009collective,martin2013collective,norcia2018collective,davis2020collective,bilitewski2022dipoleLargeSpin}. These collective spins can then be coupled via XXZ anisotropic spin interactions generated via the interplay of superexchange, spin-orbit coupling ~\cite{dalibard2011soc,atala2014soc,mancini2015soc,stuhl2015soc,kolkowitz2017soc,tai2017soc} and a linear tilt ~\cite{simon2011tiltedLattice,nichols2019tiltedLattice,dimitrova2020superexchange,aeppli2022tiltedLattice} (i.e. gravity). The resulting dynamical behavior can remain fully collective across a wide range of anisotropies, including those for which $|\Delta-1|$ is not small, and even outside the easy-plane regime $|\Delta| > 1$, allowing for scalable squeezing generation on faster timescales. Vacancies in the initial loadout are also not detrimental since they just effectively reduce the size of the collective spins coupled by the XXZ interactions.

In Section~\ref{sec_Spin} we show how nearest-neighbour spin-1/2 XXZ interactions with mixed anisotropy can be used to emulate a 1D large-spin XXZ model. We demonstrate that this model generates squeezing that scales with number of spins $N$ as $N^{-2/3}$, equivalent to the paradigmatic one-axis twisting model~\cite{kitagawa1993squeezing}. A Holstein-Primakoff approximation is used to determine the regime of validity of the large-spin model, and to provide analytical results for the system's short time behavior. In Section~\ref{sec_Implementation} we describe a detailed protocol for implementing the system with 3D optical lattices using the interplay of interactions, tunneling, spin-orbit coupling and a tilt. The SOC generates anisotropy by dressing the spins, while the tilt enables further tunability of interactions without requiring experimentally challenging tasks like precisely tuned laser angles. We also discuss the system's capacity to reverse the generated squeezing by flipping the sign of the squeezing interaction, which can be used to simplify the required noise resolution  in quantum-enhanced phase estimation in real metrological protocols~\cite{davis2016unsqueezing,schulte2020unsqueezing}. The effect of non-unit filling fraction is also studied, and demonstrated to not disrupt the squeezing. Section~\ref{sec_Outlook} provides conclusions and outlook.

\section{Spin-1/2 model description}
\label{sec_Spin}

\subsection{Large spin mapping}

We consider the situation where spin-1/2 atoms loaded into a three-dimensional lattice interact via a nearest-neighbour XXZ-type model. We assume there are $L_{\nu}$ lattice sites along each direction $ \nu \in \{\mathrm{X}, \mathrm{Y}, \mathrm{Z}\}$, with a total of $N = L_{\mathrm{X}} L_{\mathrm{Y}} L_{\mathrm{Z}}$ spins. The Hamiltonian reads,
\begin{equation}
\hat{H}_{\mathrm{S}} = \sum_{\nu = \mathrm{X},\mathrm{Y},\mathrm{Z}} V_{\nu} \sum_{\vec{r}} \left[\hat{s}_{\vec{r}}^{x} \hat{s}_{\vec{r} + \vec{\nu}}^{x}+\hat{s}_{\vec{r}}^{y} \hat{s}_{\vec{r} + \vec{\nu}}^{y} + \Delta_{\nu} \hat{s}_{\vec{r}}^{z}\hat{s}_{\vec{r}+\vec{\nu}}^{z}\right],
\end{equation}
where $\hat{s}_{\vec{r}}^{\alpha}$ is the spin-1/2 operator  $\alpha \in \{x,y,z\}$ at site $\vec{r} = (r_{\mathrm{X}},r_{\mathrm{Y}},r_{\mathrm{Z}})$ satisfying commutation relations $[\hat{s}_{\vec{r}}^{\alpha},\hat{s}_{\vec{r}}^{\beta}]= i \epsilon_{\alpha\beta\gamma}\hat{s}_{\vec{r}}^{\gamma}$, and $\vec{\nu}$ is the unit vector along direction $\nu$, i.e. $\vec{\mathrm{X}} = (1,0,0)$. The coefficients $V_{\nu}$ and $\Delta_{\nu}$ set the respective strength and anisotropy of interactions along $\nu$. We use periodic boundary conditions along all directions.

Our regime of interest is the situation where the $\mathrm{X}$, $\mathrm{Y}$ directions are 
isotropic, but anisotropy is present along the $\mathrm{Z}$ direction:
\begin{equation}
\Delta_{\mathrm{X}} = \Delta_{\mathrm{Y}} = 1, \quad \Delta_{\mathrm{Z}} \neq 1.
\end{equation}
We also set the $\mathrm{X}$, $\mathrm{Y}$ isotropic interaction strengths equal, $V_{\mathrm{X}}=V_{\mathrm{Y}} \equiv V_{\mathrm{XY}}$ for simplicity, although this is not mandatory for our protocol.

\begin{figure}[h]
\includegraphics[width=0.45\textwidth]{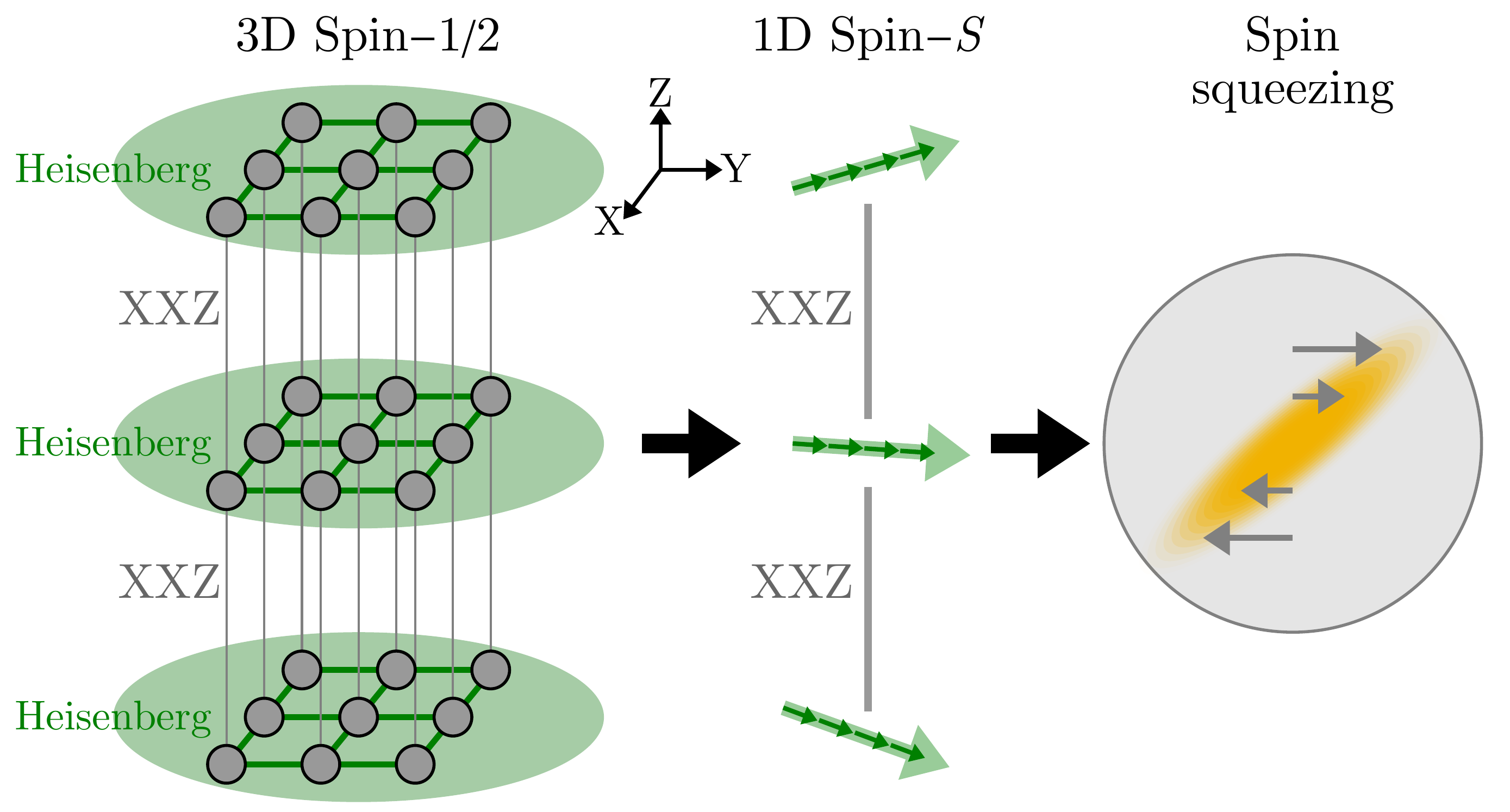}
\centering
\caption{Schematic: We consider a 3D lattice where each $\mathrm{X}-\mathrm{Y}$ plane is coupled by isotropic Heisenberg interactions while the $\mathrm{Z}$ direction has XXZ spin couplings. When the in-plane interactions are  much stronger than the inter-plane ones, an initially-prepared spin-coherent state is energetically locked by the ferromagnetic in plane interactions into a large collective spin in each plane. The XXZ interactions couple these large spins and enable the generation of scalable spin squeezing.}
\label{fig_Schematic}
\end{figure}

In this regime, the spins within each $\mathrm{X}-\mathrm{Y}$ plane of the lattice interact via a Heisenberg model with a coupling strength $V_{\mathrm{XY}}$. If the system is initially prepared in a spin-coherent collective state with all spins  aligned along a specific direction (in the Dicke manifold), and the Heisenberg coupling $V_{\mathrm{XY}}$ is much stronger than any anisotropic interactions along $\mathrm{Z}$, the spins of each plane will remain locked thanks to the large energy cost of flipping an individual spin within the planes. As a result, the anisotropic interactions are energetically  projected into the Dicke manifolds of each plane. The 3D spin-1/2 model becomes a 1D spin-$S$ model, with each $\mathrm{X}-\mathrm{Y}$ plane acting as a large spin of size $S = L_{\mathrm{X}} L_{\mathrm{Y}}/2$. Fig.~\ref{fig_Schematic} depicts this mapping.

Mathematically, we define large-spin operators for each plane indexed by their $\mathrm{Z}$ position $r_{\mathrm{Z}}$,
\begin{equation}
\hat{S}_{r_{\mathrm{Z}}}^{\alpha} = \sum_{r_{\mathrm{X}},r_{\mathrm{Y}}} \hat{s}_{\vec{r}}^{\alpha},
\end{equation}
where $\hat{S}_{r_{\mathrm{Z}}}^{\alpha}$ still obeys commutation relations $[\hat{S}_{r_{\mathrm{Z}}}^{\alpha},\hat{S}_{r_{\mathrm{Z}}}^{\beta}]= i \epsilon_{\alpha\beta\gamma}\hat{S}_{r_{\mathrm{Z}}}^{\gamma}$, although the operators are of spin size $S$ rather than 1/2. We assume that the underlying spin-1/2 operators can be written in terms of these large-spin operators as,
\begin{equation}
\label{eq_SpinToLargeSpinOps}
\hat{s}_{\vec{r}}^{\alpha} = \frac{1}{L_{\mathrm{X}}L_{\mathrm{Y}}} \hat{S}_{r_{\mathrm{Z}}}^{\alpha},
\end{equation}
and thus effectively project the spin-1/2 degrees of freedom into the Dicke manifold of each $\mathrm{X}$-$\mathrm{Y}$ plane. The spin Hamiltonian then reads,
\begin{equation}
\label{eq_LargeSpinModel}
\begin{aligned}
\hat{H}_{\mathrm{LS}} &= \frac{1}{L_{\mathrm{X}}L_{\mathrm{Y}}}\sum_{r_{\mathrm{Z}}} \left[V_{\mathrm{Z}}\vec{S}_{r_{\mathrm{Z}}}\cdot\vec{S}_{r_{\mathrm{Z}}+1}+\chi\hat{S}_{r_{\mathrm{Z}}}^{z}\hat{S}_{r_{\mathrm{Z}}+1}^{z}\right]\\
&+\frac{V_{\mathrm{X}}+V_{\mathrm{Y}}}{L_{\mathrm{X}}L_{\mathrm{Y}}}\sum_{r_{\mathrm{Z}}}\vec{S}_{r_{\mathrm{Z}}}\cdot \vec{S}_{r_{\mathrm{Z}}},
\end{aligned}
\end{equation}
where $\vec{S}_{r_{\mathrm{Z}}} = (\hat{S}_{r_{\mathrm{Z}}}^{x},\hat{S}_{r_{\mathrm{Z}}}^{y},\hat{S}_{r_{\mathrm{Z}}}^{z})$, and $\chi$ is the anisotropic portion of the interactions that drives squeezing dynamics,
\begin{equation}
\chi=V_{\mathrm{Z}}(\Delta_{\mathrm{Z}}-1).
\end{equation}
The terms on the second line of Eq.~\eqref{eq_LargeSpinModel} are the in-plane Heisenberg interactions, which do not affect the dynamics of relevant spin observables and are omitted from $\hat{H}_{\mathrm{LS}}$ hereafter.

\begin{figure}[h]
\includegraphics[width=0.43\textwidth]{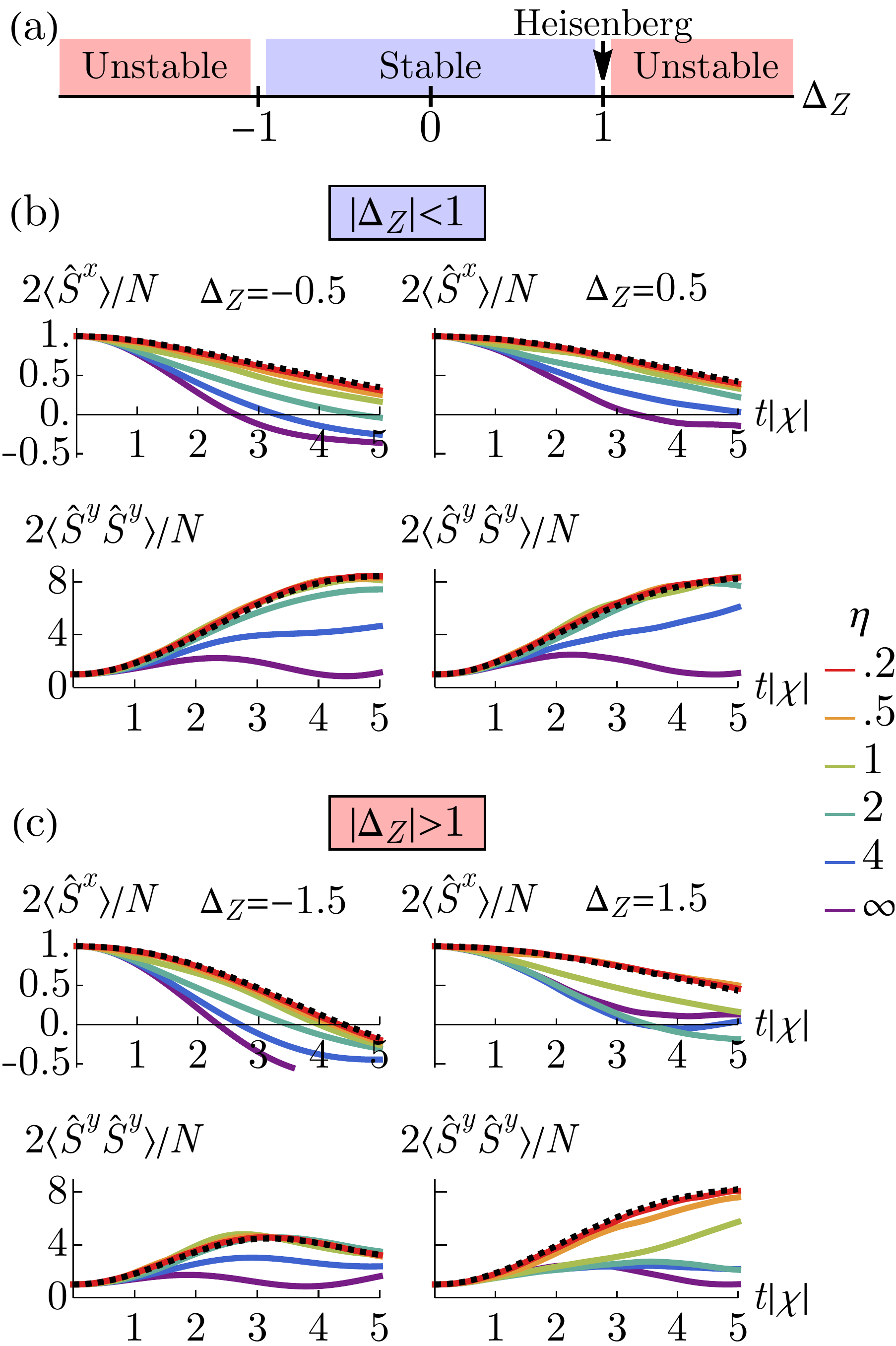}
\centering
\caption{
(a) Anisotropy of the spin model along $\mathrm{Z}$. The Heisenberg point is at $\Delta_{\mathrm{Z}} = 1$. Bogoliubov modes in a Holstein-Primakoff mapping are parametrically stable for $|\Delta_{\mathrm{Z}}|<1$.
(b) Time evolution of spin observables $\langle\hat{S}^{x}\rangle$ and $\langle\hat{S}^{y}\hat{S}^{y}\rangle$ for the spin-1/2 model $\hat{H}_{\mathrm{S}}$ (colored lines) and the large-spin model $\hat{H}_{\mathrm{LS}}$ (black dashed line). Anisotropies in the stable regime are chosen, with values $\Delta_{\mathrm{Z}}=-0.5$ and $0.5$ (left and right panels). The ratio of the in-plane to anisotropic out-of-plane interactions is $\eta = |\chi/V_{\mathrm{XY}}|$ with $\chi=V_{\mathrm{Z}}(\Delta_{\mathrm{Z}}-1)$. The system size is $L_{\mathrm{X}} \times L_{\mathrm{Y}} \times L_{\mathrm{Z}}=4\times 1\times 4$. (c) Same comparisons as panel (b) but for unstable anisotropies $\Delta_{\mathrm{Z}} = -1.5$ and $\Delta_{\mathrm{Z}} = 1.5$.}
\label{fig_Spin12ToLargeSpin}
\end{figure}

The validity of the mapping from $\hat{H}_{\mathrm{S}}$ to $\hat{H}_{\mathrm{LS}}$, i.e. the projection of the spin operators in Eq.~\eqref{eq_SpinToLargeSpinOps}, relies upon the system not creating in-plane spin-wave excitations, thus allowing each large spin to stay collective. A detailed analysis of the validity using a Holstein-Primakoff approximation is provided in Appendix~\ref{app_LargeSpin}. In the regime $|\Delta_{\mathrm{Z}}|<1$ the system has no unstable Bogoliubov modes, as depicted in Fig.~\ref{fig_Spin12ToLargeSpin}(a). In this regime we show analytically that for an initial state that is fully collective across all spins (not just individual planes), the number of non-collective in-plane excitations is bounded by $\sim \eta N$, where $\eta$ is a small parameter characterizing the strength of out-of-plane anisotropic interactions to the in-plane ones,
\begin{equation}
\eta \equiv \frac{|\chi|}{|V_{\mathrm{XY}}|}.
\end{equation}
When the bounded number of non-collective excitations is small compared to $N$, which we show to hold for $\eta \ll 1$ via the Holstein-Primakoff approximation in Appendix~\ref{app_LargeSpin}, the time evolution of all extensive collective-spin observables relevant to squeezing is captured by $\hat{H}_{\mathrm{LS}}$.

We test the validity of the mapping by comparing time evolution under $\hat{H}_{\mathrm{S}}$ and $\hat{H}_{\mathrm{LS}}$ using exact numerical integration of the Schr\"{o}dinger equation. The initial state for the evolution is all spin-1/2s pointing along the $+x$ direction of the Bloch sphere,
\begin{equation}
\ket{\psi_0} = \bigotimes_{\vec{r}} \frac{1}{\sqrt{2}} \left(\ket{\uparrow}_{\vec{r}} + \ket{\downarrow}_{\vec{r}}\right).
\end{equation}
For the large-spin model this state is,
\begin{equation}
\ket{\psi_0} = \bigotimes_{r_{\mathrm{Z}}}e^{-i \frac{\pi}{2}\hat{S}_{r_{\mathrm{Z}}}^{y}}\ket{S,-S}_{r_{\mathrm{Z}}},
\end{equation}
where $\ket{S,M}_{r_{\mathrm{Z}}}$ is the Dicke state of the large spin at plane $r_{\mathrm{Z}}$ with total angular momentum $S$ and projection $M \in \{-S \dots S\}$.

Fig.~\ref{fig_Spin12ToLargeSpin}(b) plots the time evolution of sample collective spin observables relevant to spin squeezing. We study the single-particle observable $\langle\hat{S}^{x}\rangle$ (where $\hat{S}^{\alpha}=\sum_{r_{\mathrm{Z}}}\hat{S}_{r_{\mathrm{z}}}^{\alpha} = \sum_{\vec{r}}\hat{s}_{\vec{r}}^{\alpha}$) which characterizes the spin contrast of the system and the length of the collective spin vector. We also look at spin correlations such as $\langle\hat{S}^{y}\hat{S}^{y} \rangle$, which is one of the correlators relevant to squeezing from this initial state that characterizes the minimum noise distribution of the collective spin (see Appendix~\ref{app_LargeSpin} for details). These spin observables are studied for a few anisotropy values within the stable regime $|\Delta_{\mathrm{Z}}|<1$. The rate of the dynamics is set by the anisotropic interaction strength $\chi$. The large-spin model reproduces the behavior of the spin-1/2 model for $\eta \ll 1$. These numerical benchmarks are done for a 2D system with $L_{\mathrm{Y}}=1$ rather than full 3D due to the limited size of numerically tractable systems, but we expect that the protocols will be even more effective in 3D where $S$ can be made much larger.

As an interesting extension, in Fig.~\ref{fig_Spin12ToLargeSpin}(c) we also make the same comparison for anisotropies in the unstable regime $|\Delta_{\mathrm{Z}}|>1$. Despite the exponential growth of certain non-collective Bogoliubov modes in this regime, we still find good agreement when $\eta \ll 1$ out to long timescales $t|\chi| \gg 1$. Qualitatively, this extended agreement holds because the strong in-plane interactions restrict the number of unstable modes and reduce their contributions to the non-collective dynamics. While it is more difficult to extrapolate our analytic bounds (Appendix~\ref{app_LargeSpin}) to this regime, we will qualitatively find further on that scalable squeezing emerges in the unstable regime nonetheless.

\subsection{Spin squeezing}

We now study the squeezing properties of the large-spin model. The spin squeezing is defined as,
\begin{equation}
\zeta^2 = N\text{min}_{\theta} \frac{\langle \Delta \vec{S}_{\perp \theta}\rangle}{|\langle\vec{S}\rangle|^2},
\end{equation}
where $\vec{S} = (\hat{S}^{x},\hat{S}^{y},\hat{S}^{z})$ is the collective spin of the full system, and $\Delta \vec{S}_{\perp \theta}$ is the variance of the collective spin along an axis perpendicular to its direction, parametrized by an angle $\theta \in [0,2\pi]$. The value of $\zeta^2$ characterizes the improvement in sensitivity over a spin coherent state.

\begin{figure}[h]
\includegraphics[width=0.46\textwidth]{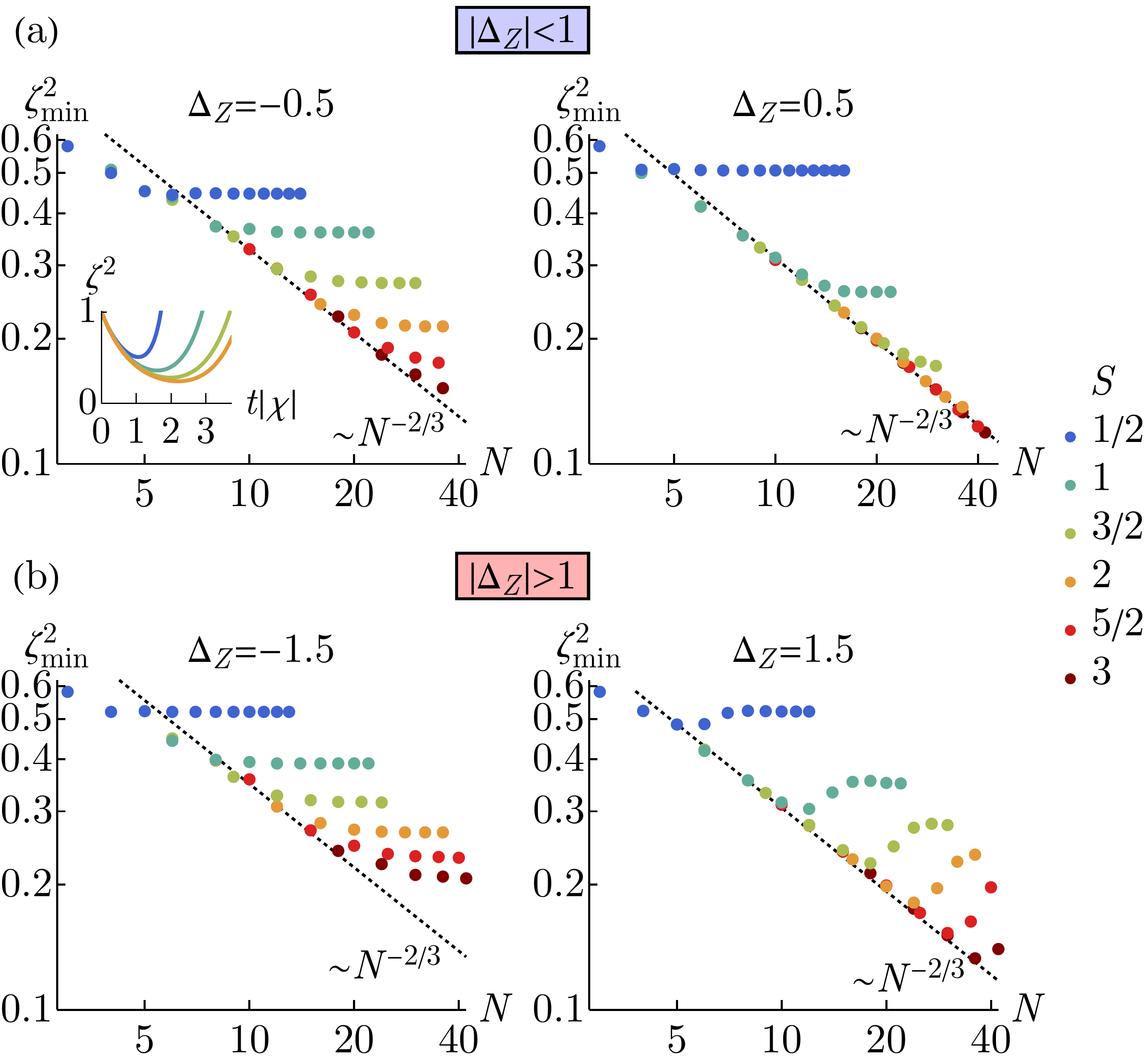}
\centering
\caption{(a) Optimal squeezing $\zeta_{\mathrm{min}}^2 = \text{min}_{t} \zeta^2 (t)$ generated by the large-spin model $\hat{H}_{\mathrm{LS}}$ in the stable regime for anisotropies $\Delta_{\mathrm{Z}}= -0.5$ and $\Delta_{\mathrm{Z}}=0.5$ (left and right panels), in terms of total number of spin-1/2 constituents $N = L_{\mathrm{X}} L_{\mathrm{Y}} L_{\mathrm{Z}}$. Different colors indicate different large spin sizes $S = L_{\mathrm{X}} L_{\mathrm{Y}}/2$; for a fixed $S$, total $N$ is varied by changing the number of planes $L_{\mathrm{Z}}$. The squeezing is compared to the ideal one-axis twisting scaling of $\sim N^{-2/3}$; larger spin sizes follow this scaling for larger total spin number $N$. The inset shows sample transient evolution of the squeezing for roughly fixed system sizes of $N \approx 16$. (b) Optimal squeezing in the unstable regime for anisotropies $\Delta_{\mathrm{Z}} = -1.5$ and $\Delta_{\mathrm{Z}}= 1.5$.}
\label{fig_LargeSpinSqueezingScaling}
\end{figure}

In Fig.~\ref{fig_LargeSpinSqueezingScaling} we show the optimal squeezing attained by the large-spin model for different spin sizes $S$. We find that even for anisotropies far from the Heisenberg point the squeezing scales with $N$ for sufficiently large $S$. This scalable squeezing is generated because the system behaves collectively, not just within individual planes but in its entirety. We can further project the large-spin model into the Dicke manifold, yielding a one-axis twisting (OAT) Hamiltonian:
\begin{equation}
    \hat{H}_{\mathrm{LS}} \approx \frac{V_{\mathrm{Z}}}{N} \vec{S} \cdot \vec{S} + \frac{\chi}{N}\hat{S}^{z}\hat{S}^{z}.
\end{equation}
The $\vec{S}\cdot\vec{S}$ term is a constant for a collective initial state and does not affect the dynamics. The OAT model generates squeezing that scales as $\sim N^{-2/3}$, and the large-spin model agrees with this scaling for all the system sizes shown.

For general spin-1/2 XXZ systems, the validity of the OAT model improves in higher dimensions and for anisotropies closer to the Heisenberg point~\cite{perlin2020spin, comparin2022towerOfStates, frerot2017squeezingXYferromagnet, block2023squeezingScaling}.
Our approach instead exploits the dimensionality of the system, using strong in-plane interactions to maintain the ferromagnetic order while still having a broad range of viable anisotropies along the out-of-plane direction. In particular, being closer to $\Delta_{\mathrm{Z}}=1$ yields a better overall prefactor for the squeezing at the cost of slower timescales, but the scaling with $N$ remains the same. When the system stays collective, the time needed to reach the optimum squeezing scales as $t_{\mathrm{min}} |\chi|\sim N^{1/3}$.

As an aside, we comment that the specific scaling exponent of $-2/3$ for the optimal squeezing is not guaranteed to hold in the $N\to\infty$ limit. Recent studies suggest that spin-1/2 models with anisotropy along all directions can exhibit different exponents for very large systems $N\gg 10^3$~\cite{block2023squeezingScaling}. The effects of boundary conditions may also not be negligible even in the thermodynamic limit~\cite{tanausu2023squeezingScaling}. Regardless, even the small systems with $N\approx 50$ spins used in our simulations can generate squeezing in excess of 10 dB, and this value will continue to improve with increasing $N$.

\begin{figure}[h]
\includegraphics[width=0.48\textwidth]{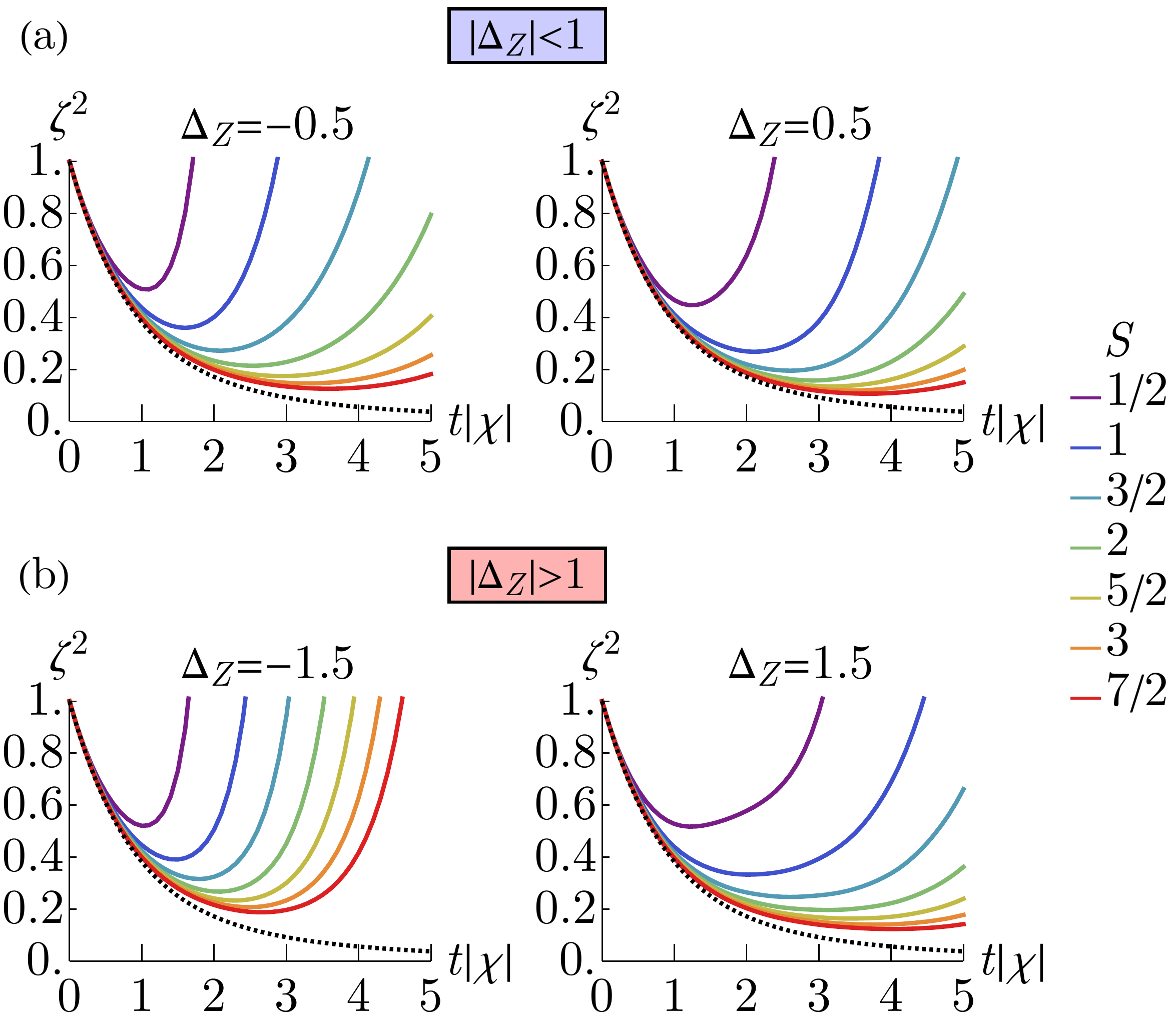}
\centering
\caption{(a) Time-evolution of squeezing generated by the large-spin model for increasing spin size $S$ and a fixed number of large spins $L_{\mathrm{Z}}=7$, in stable regime for anisotropies $\Delta_{\mathrm{Z}}=-0.5$ and $\Delta_{\mathrm{Z}} = 0.5$ (left and right panels). The black dashed line is the Holstein-Primakoff prediction from Eq.~\eqref{eq_HolsteinSqueezing}. (b) Time-evolution of squeezing in the unstable regime for anisotropies $\Delta_{\mathrm{Z}}=-1.5$ and $\Delta_{\mathrm{Z}} = 1.5$.}
\label{fig_Holstein}
\end{figure}

We can also obtain some analytic results by studying the dynamics of the zero-quasimomentum mode $\vec{k}=0$ under a Holstein-Primakoff approximation (Appendix~\ref{app_LargeSpin}), which is often sufficient to capture squeezing properties~\cite{roscilde2023rotorSqueezing}. Solving the time-evolution of this mode explicitly, we find the following analytic expression for the squeezing,
\begin{equation}
\label{eq_HolsteinSqueezing}
\zeta^2 \approx \frac{1}{\left(\sqrt{1+\frac{\chi^2 t^2}{4}}+ \frac{|\chi| t}{2}\right)^2} + \mathcal{O}\left(\frac{1}{S}\right),
\end{equation}
which we confirm to be valid at short times in Fig.~\ref{fig_Holstein}. The validity improves with larger spin size $S$. The growth of the time needed for optimal squeezing is characteristic of the local nature of the interactions. More time is needed for entanglement to spread through the system and build up to a higher value. The formula does not give the correct scaling of the optimal squeezing with $N$, for which higher-order corrections are needed. As commented previously, however, the exact scaling in the $N \to \infty$ limit can be subtle and we leave more careful analysis to future work.

\section{Implementation with 3D optical lattices}
\label{sec_Implementation}
\subsection{Lattice model}

We now provide an in-depth discussion on how to implement the large-spin model in optical lattice  arrays. Our treatment focuses on fermionic atoms, but extends naturally to bosons.

We consider atoms with two internal states $\sigma \in \{g,e\}$ loaded into the lowest motional band of a 3D optical lattice. The system has $L_{\mathrm{X}} \times L_{\mathrm{Y}} \times L_{\mathrm{Z}}$ lattice sites populated by a total of $N$ atoms. The atoms feel a linear potential such as gravity along the $\mathrm{Z}$ direction, and are illuminated by an additional driving laser that induces spin-flips while transferring momentum to the atoms; this laser is also pointed along $\mathrm{Z}$. The Hamiltonian describing this system is a Hubbard-type model,
\begin{equation}
\hat{H} = \hat{H}_{\mathrm{Tunneling}} + \hat{H}_{\mathrm{Interaction}} + \hat{H}_{\mathrm{Gravity}} + \hat{H}_{\mathrm{Drive}}.
\end{equation}
Here $\hat{H}_{\mathrm{Tunneling}}$ is the nearest-neighbour tunneling of atoms,
\begin{equation}
\hat{H}_{\mathrm{Tunneling}} = -\sum_{\nu=\mathrm{X},\mathrm{Y},\mathrm{Z}} J_{\nu} \sum_{\vec{r},\sigma} \left(\hat{c}_{\vec{r},\sigma}^{\dagger}\hat{c}_{\vec{r}+ \vec{\nu},\sigma}+h.c.\right),
\end{equation}
where $J_{\nu}$ is the tunneling rate along $\nu$, and $\hat{c}_{\vec{r},\sigma}$ annihilates an atom of spin $\sigma$ at lattice site index $\vec{r} = (r_{\mathrm{X}}, r_{\mathrm{Y}}, r_{\mathrm{Z}})$ as before.

The interaction Hamiltonian is an on-site Hubbard  term that accounts for collisions between the atoms,
\begin{equation}
\hat{H}_{\mathrm{Interaction}} = U\sum_{\vec{r}}\hat{n}_{\vec{r},g}\hat{n}_{\vec{r},e},
\end{equation}
where $\hat{n}_{\vec{r},\sigma} = \hat{c}_{\vec{r},\sigma}^{\dagger}\hat{c}_{\vec{r},\sigma}$ and $U$ is the Hubbard interaction strength proportional to the singlet $e$-$g$ scattering length between the atoms.

The linear potential term is,
\begin{equation}
\hat{H}_{\mathrm{Gravity}} = G \sum_{\vec{r}} r_{\mathrm{Z}} \left(\hat{n}_{\vec{r},g} + \hat{n}_{\vec{r},e}\right),
\end{equation}
with $G$ the potential difference per lattice site. This term arises naturally due to gravity, with a typical strength $G=mga$ for atomic mass $m$, gravitational acceleration $g$ and lattice spacing $a$. Such a term can also be engineered through the use of e.g. Stark shifts, or accelerated lattices~\cite{zheng2022accelerated}.

The final term is an applied laser drive,
\begin{equation}
\hat{H}_{\mathrm{Drive}} = \frac{\Omega}{2} \sum_{\vec{r}} \left(e^{i a\vec{k}_{c} \cdot \vec{r}} \hat{c}_{\vec{r},e}^{\dagger}\hat{c}_{\vec{r},g} + h.c.\right),
\end{equation}
where $\Omega$ is the laser Rabi frequency, and $ a\vec{k}_c \cdot \vec{r}$ is the laser coupling spatial Peierls phase with $\vec{k}_c$ the laser wavevector and $a \vec{r}$ the lattice position. Since the laser is pointed along $\mathrm{Z}$ we write $\vec{k}_c = (0,0,|\vec{k}_c|)$, for which the phase takes the form,
\begin{equation}
a\vec{k}_c \cdot \vec{r} =  r_{\mathrm{Z}} \phi,
\end{equation}
with $\phi= a \vec{k}_{c}\cdot \vec{\mathrm{Z}}$ the differential phase per lattice site along $\mathrm{Z}$. This phase generates spin-orbit coupling (SOC)~\cite{dalibard2011soc}, whose effect has been observed in several optical lattice experiments~\cite{atala2014soc,mancini2015soc,stuhl2015soc,kolkowitz2017soc,tai2017soc,bromley2018soc}. In our case the SOC phase will induce anisotropy in the effective spin interactions between the atoms.

Unlike the preceding section we consider open boundary conditions along all directions to match realistic experimental conditions.

\subsection{Dressing}
\label{sec_Dressing}

We assume that the driving laser stays on during the system dynamics. The presence of the drive dresses the spin states of the atoms. These dressed states, labeled as $\tilde{\sigma} \in \{\uparrow,\downarrow \}$, correspond to the on-site single particle eigenstates of $\hat{H}_{\mathrm{Drive}}$, with energy $\{+\frac{\Omega}{2}, -\frac{\Omega}{2} \}$ respectively. The fermionic annihilation operators associated with these eigenstates are,
\begin{equation}
\begin{aligned}
\hat{a}_{\vec{r},\uparrow} &= \frac{1}{\sqrt{2}} \left(e^{i r_{\mathrm{Z}} \phi/2} \hat{c}_{\vec{r},g} + e^{- i r_{\mathrm{Z}} \phi/2} \hat{c}_{\vec{r},e}\right),\\
\hat{a}_{\vec{r},\downarrow} &= \frac{1}{\sqrt{2}} \left(e^{i r_{\mathrm{Z}} \phi/2} \hat{c}_{\vec{r},g} - e^{-i r_{\mathrm{Z}} \phi/2} \hat{c}_{\vec{r},e}\right),\\
\end{aligned}
\end{equation}
where $\hat{a}_{\vec{r},\tilde{\sigma}}$ annihilates an atom in dressed state $\tilde{\sigma}$ on site $\vec{r}$. The clock laser is thus diagonal in the dressed basis,
\begin{equation}
\hat{H}_{\mathrm{Drive}} = \frac{\Omega}{2} \sum_{\vec{r}} \left(\hat{n}_{\vec{r},\uparrow} - \hat{n}_{\vec{r},\downarrow}\right).
\end{equation}
The interactions and gravity retain their form,
\begin{equation}
\begin{aligned}
\hat{H}_{\mathrm{Interaction}} &= U \sum_{\vec{r}} \hat{n}_{\vec{r},\uparrow} \hat{n}_{\vec{r},\downarrow},\\
\hat{H}_{\mathrm{Gravity}}&= G \sum_{\vec{r}} r_Z \left(\hat{n}_{\vec{r},\uparrow} + \hat{n}_{\vec{r},\downarrow}\right).
\end{aligned}
\end{equation}
The tunneling Hamiltonian takes the form of,
\begin{equation}
\begin{aligned}
&\hat{H}_{\mathrm{Tunneling}}= -\sum_{\nu=\mathrm{X},\mathrm{Y}} J_{\nu}\sum_{\vec{r},\tilde{\sigma}} \left(\hat{a}_{\vec{r},\tilde{\sigma}}^{\dagger} \hat{a}_{\vec{r}+ \vec{\nu},\tilde{\sigma}} + h.c.\right)\\
&-J_{\mathrm{Z}}\sum_{\vec{r},\tilde{\sigma}}\bigg[\cos\left(\frac{\phi}{2}\right)\hat{a}_{\vec{r},\tilde{\sigma}}^{\dagger}\hat{a}_{\vec{r}+\vec{\mathrm{Z}},\tilde{\sigma}} -i \sin\left(\frac{\phi}{2}\right)\hat{a}_{\vec{r},\tilde{\sigma}}^{\dagger}\hat{a}_{\vec{r}+\vec{\mathrm{Z}},\tilde{\sigma}'}  \\
&\quad \quad + h.c. \bigg],
\end{aligned}
\end{equation}
where $\tilde{\sigma}'$ is a flipped spin, $\uparrow' = \downarrow$ and $\downarrow' = \uparrow$, since  the term proportional to $\sin\left(\frac{\phi}{2}\right)$ corresponds to hopping terms along $\mathrm{Z}$ that also flip the spin of the dressed states. For $\phi = 0$ we only have dressed-spin-conserving tunneling, whereas for $\phi = \pi$ the atoms must flip dressed spin when they hop along $\mathrm{Z}$.

Atoms can be converted from the dressed basis $\{\uparrow,\downarrow\}$ to the original spin basis $\{g,e\}$ or vice versa with a laser pulse. For example, applying a $\pi/2$ pulse via the driving laser with a phase shift of $\pi/2$ relative to its original frame will convert a state of all atoms in $g$ to all in $\uparrow$:
\begin{equation}
\label{eq_DressedPulse}
\exp \left(-i \frac{\pi}{2\Omega}\hat{H}_{\mathrm{Drive}}^{\mathrm{shift}}\right)\prod_{\vec{r}} \hat{c}_{\vec{r},g}^{\dagger}\ket{0}=\prod_{\vec{r}}\hat{a}_{\vec{r},\uparrow}^{\dagger}\ket{0},
\end{equation}
where $\ket{0}$ is the vacuum and $\hat{H}_{\mathrm{Drive}}^{\mathrm{shift}}$ is the laser drive Hamiltonian with a global $\pi/2$ phase shift.
\begin{equation}
\hat{H}_{\mathrm{Drive}}^{\mathrm{shift}}=\frac{\Omega}{2}\sum_{\vec{r}}\left(e^{i(\vec{k}_{c}\cdot \vec{r} + \pi/2)}\hat{c}_{\vec{r},e}^{\dagger}\hat{c}_{\vec{r},g}+h.c.\right).
\end{equation}
The same pulse will rotate all $e$ into all $\uparrow$. One can use such a pulse to rotate a squeezed state from the dressed basis into the original spin basis, whereupon it can undergo Ramsey precession for metrological purposes. One can also prepare/unprepare dressed states by adiabatically ramping the detuning of the laser; see Ref.~\cite{mamaev2022resonant} for further discussion.

For clarity, before delving into details of the spin interactions we outline how a squeezing operation would proceed. We start with a superposition of $\uparrow$ and $\downarrow$ dressed states along the $+x$ direction of the dressed Bloch sphere, which will undergo squeezing dynamics. Conveniently, this state corresponds to all atoms in the non-dressed ground spin state $g$,
\begin{equation}
    \ket{\psi_0} = \prod_{\vec{r}} \frac{1}{\sqrt{2}} \left(\hat{a}_{\vec{r},\uparrow}^{\dagger} + \hat{a}_{\vec{r},\downarrow}^{\dagger}\right)\ket{0} = \prod_{\vec{r}}\hat{c}_{\vec{r},g}^{\dagger}\ket{0}.
\end{equation}
This state is prepared via standard cooling protocols, the lattice is made shallow enough to allow tunneling, and the drive Hamiltonian $\hat{H}_{\mathrm{Drive}}$ is turned on. The system is allowed to evolve under the laser drive on timescales $|\chi| t \sim N^{1/3}$, which will generate a squeezed state in the dressed basis $\{\downarrow,\uparrow\}$.

The squeezed state will generally not have its reduced noise distribution aligned along the equator of the collective Bloch sphere (which is the phase sensitive direction). However, one may rotate the squeezed state about its spin direction axis with a unitary operation $e^{-i \theta \hat{S}^{x}}$,  choosing a $\theta$ that aligns the reduced noise distribution along the equator. This rotation is straightforward to implement because it corresponds to a laser detuning in the original spin basis: $\hat{S}^{x}= \frac{1}{2}\sum_{\vec{r}}\left(\hat{n}_{\vec{r},e}-\hat{n}_{\vec{r},g}\right)$.

Once the squeezed state is aligned to be phase-sensitive along the equator of the dressed Bloch sphere, it can be rotated into the original spin basis $\{g,e\}$ via the pulse in Eq.~\eqref{eq_DressedPulse}. This pulse can be implemented with the same driving laser used during the squeezing generation by skipping its phase ahead by $\pi/2$, which turns $\hat{H}_{\mathrm{Drive}}$ into $\hat{H}_{\mathrm{Drive}}^{\mathrm{shift}}$. After this phase skip the drive kept on for a time $t\Omega = \pi/2$ (assumed much faster than lattice dynamics), implementing the pulse. The resulting squeezed state can now be directly used for conventional Ramsey precession in a clock operation.

\subsection{Superexchange interactions}

We assume an initial filling fraction of one atom per site, and strong on-site interactions compared to tunneling rates $U \gg J_{\nu}$. The low-energy behavior physics can be described with a superexchange spin-1/2 model. The spin operators are defined in terms of the dressed states,
\begin{equation}
\label{eq_SpinOperatorDefinitions}
\begin{aligned}
\hat{s}_{\vec{r}}^{x} &= \frac{1}{2}\left(\hat{a}_{\vec{r},\uparrow}^{\dagger}\hat{a}_{\vec{r},\downarrow} + h.c.\right),\\
\hat{s}_{\vec{r}}^{y} &= -\frac{i}{2}(\hat{a}_{\vec{r},\uparrow}^{\dagger}\hat{a}_{\vec{r},\downarrow} - h.c.),\\
\hat{s}_{\vec{r}}^{z} &= \frac{1}{2}\left(\hat{n}_{\vec{r},\uparrow} - \hat{n}_{\vec{r},\downarrow}\right).\\
\end{aligned}
\end{equation}
We obtain the superexchange interactions with a  Schrieffer-Wolff transformation up to second order in  perturbation theory~\cite{bravyi2011schriefferWolff}; see Appendix~\ref{app_Superexchange} for the derivation. The interactions are isotropic along $\mathrm{X}$, $\mathrm{Y}$, but anisotropic along $\mathrm{Z}$ due to the dressing,
\begin{equation}
\begin{aligned}
\hat{H}_{\mathrm{Spin}} = \frac{4J_{\mathrm{X}}^2}{U} &\sum_{\vec{r}} \vec{s}_{\vec{r}} \cdot \vec{s}_{\vec{r} + \vec{\mathrm{X}}}+\frac{4J_{\mathrm{Y}}^2}{U} \sum_{\vec{r}} \vec{s}_{\vec{r}} \cdot \vec{s}_{\vec{r} + \vec{\mathrm{Y}}}\\
+V_{\mathrm{Z}} &\sum_{\vec{r}} \left[\vec{s}_{\vec{r}} \cdot \vec{s}_{\vec{r}+\vec{\mathrm{Z}}} +(\Delta_{\mathrm{Z}}-1)\hat{s}_{\vec{r}}^{z} \hat{s}_{\vec{r} +\vec{\mathrm{Z}}}^{z} \right]\\
+\Omega_{\mathrm{Z}}& \sum_{\vec{r}} \left(\hat{s}_{\vec{r}}^{z} + \hat{s}_{\vec{r}+\vec{\mathrm{Z}}}^{z}\right) + \Omega \sum_{\vec{r}}\hat{s}_{\vec{r}}^{z},
\end{aligned}
\end{equation}
where $\vec{s}_{\vec{r}} = (\hat{s}^{x}_{\vec{r}},\hat{s}^{y}_{\vec{r}},\hat{s}^{z}_{\vec{r}})$, and the superexchange coefficients along $\mathrm{Z}$ are,
\begin{equation}
\begin{aligned}
\label{eq_XXZparams}
V_{\mathrm{Z}}&= \frac{4J_{\mathrm{Z}}^2 U}{U^2 - G^2}\cos^2 \left(\frac{\phi}{2}\right),\\
\Delta_{\mathrm{Z}}&=1- \frac{(U^2-G^2)(U^2-\Omega^2-G^2)}{(U^2-\Omega^2-G^2)^2-4\Omega^2G^2}\tan^2 \left(\frac{\phi}{2}\right),\\
\Omega_{\mathrm{Z}} &=-\frac{2 J_{\mathrm{Z}}^2 \Omega (U^2-\Omega^2+G^2)}{(U^2-\Omega^2+G^2)^2-4U^2 G^2}\sin^2 \left(\frac{\phi}{2}\right).
\end{aligned}
\end{equation}
In the limit where there is no SOC $\phi \to 0$, we recover a Heisenberg interaction with $\Delta_{\mathrm{Z}} = 1$ and interaction strength $V_{\mathrm{Z}} \to\frac{4 J_{\mathrm{Z}}^2 U}{U^2 - G^2}$, or just $\frac{4J_{\mathrm{Z}}^2}{U}$ in the absence of any external tilt. A similar model can also be derived for bosons (Appendix~\ref{app_Superexchange}).

\begin{figure}[h]
\includegraphics[width=0.35\textwidth]{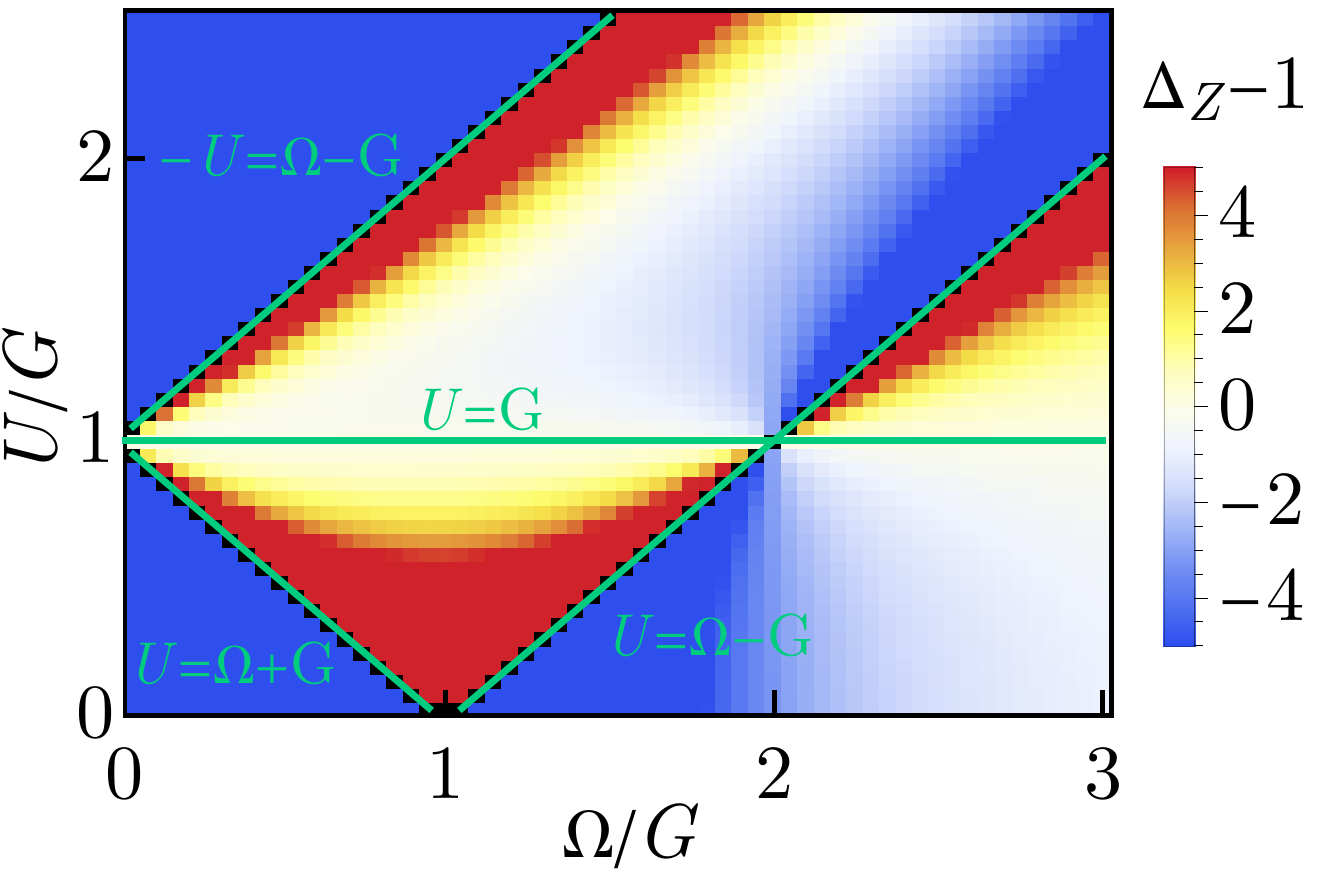}
\centering
\caption{Anisotropy of the $\mathrm{Z}$ direction superexchange interactions from Eq.~\eqref{eq_XXZparams} for an SOC phase of $\phi = 3\pi/4$, in terms of the Rabi frequency $\Omega$ and the on-site Hubbard interaction $U$. The axes are scaled in terms of the strength of gravity $G$ per site. We plot the anisotropy relative to the Heisenberg point $\Delta_{\mathrm{Z}}-1$ such that blue colors indicate $\Delta_{\mathrm{Z}}<1$ while orange colors indicate $\Delta_{\mathrm{Z}}>1$.}
\label{fig_Anisotropy}
\end{figure}

For an initially polarized state, the parameter that controls the subsequent spin dynamics is the anisotropy $\Delta_{\mathrm{Z}}$. Fig.~\ref{fig_Anisotropy} plots this anisotropy as a function of $\Omega/G$ and $U/G$ for a sample SOC phase $\phi = 3\pi/4$ (fixing $G$ to be constant). This model extends the one studied in Ref.~\cite{mamaev2021spinModel} by the inclusion of the linear potential $G$, which allows one to more easily tune $\Delta_{\mathrm{Z}}$ without having to adjust the SOC phase $\phi$ (provided that $\phi \neq 0, \pi$). This tunability is beneficial because adjusting $\phi$ in an experimental context may otherwise require changing the angle of the driving laser significantly away from the $\mathrm{Z}$ direction, which could induce non-negligible anisotropy in the $\mathrm{X}$ or $\mathrm{Y}$ directions and cause the large-spin mapping to break down. As an aside, the possibility of tuning superexchange interactions by the aid of an external tilt has already been demonstrated experimentally~\cite{dimitrova2020superexchange}.

Note that there are lines in parameter space $|U| = |\Omega \pm G|$, $|U|=|G|$,shown as teal lines, for which the denominators of the superexchange coefficients vanish. Near these lines perturbation theory breaks down because the system exhibits interaction-enabled resonant tunneling~\cite{bukov2016correlatedHopping,xu2018correlatedHopping,mamaev2022resonant}. While resonant regimes are useful for more exotic applications such as kinetically-constrained models~\cite{scherg2021kineticConstraints} or lattice gauge theory simulation~\cite{yang2020latticeGauge}, in this work we use parameters that avoid such resonances.

\subsection{Spin model benchmarking and dynamical decoupling}
\label{sec_BenchmarkingAndEcho}

\begin{figure}[h]
\includegraphics[width=0.4\textwidth]{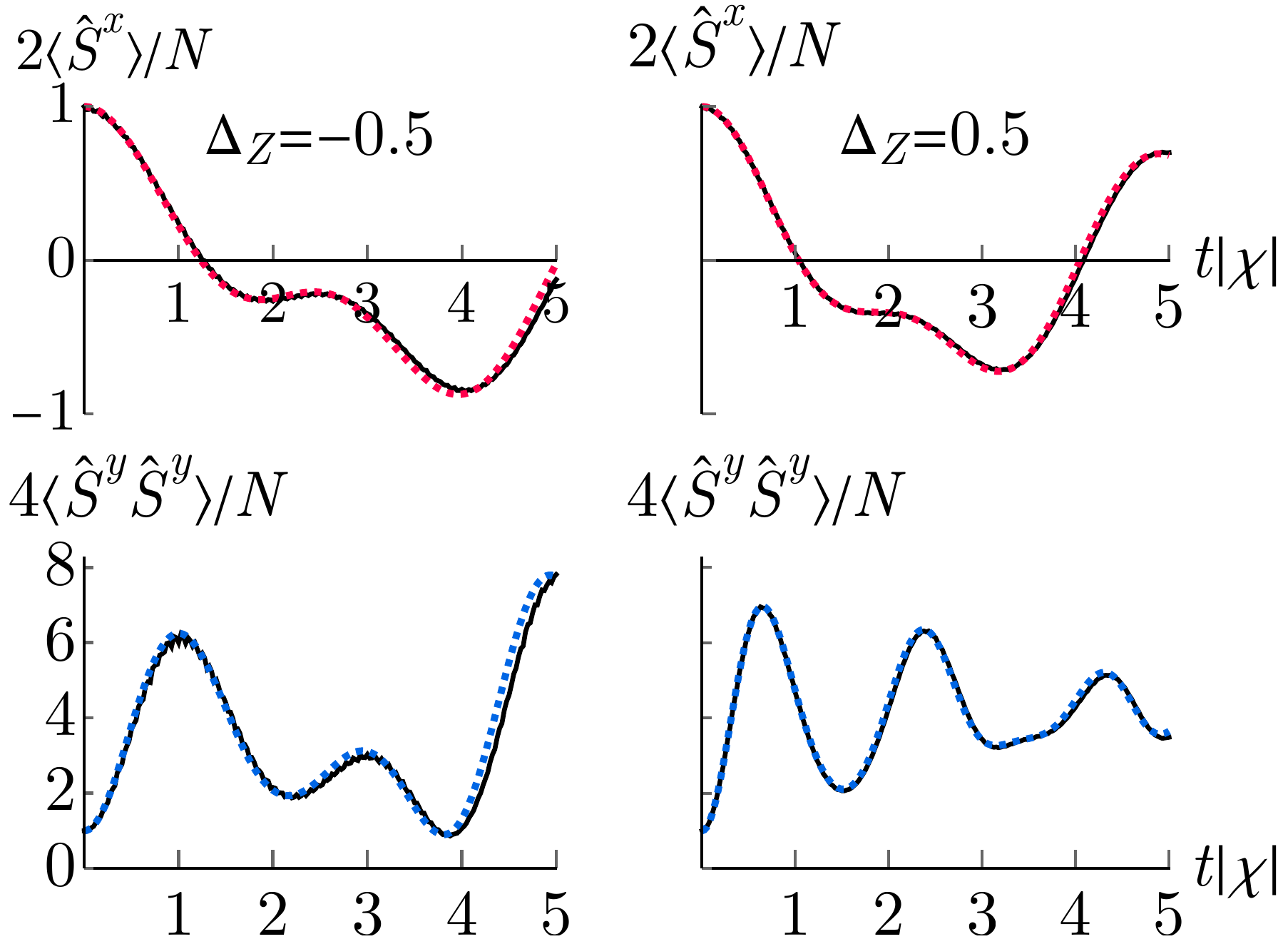}
\centering
\caption{Comparison of collective spin observables for the Hubbard model $\hat{H}$ (black lines) and the spin-1/2 superexchange model $\hat{H}_{\mathrm{SE}}$ (colored lines). The system size is $L_{\mathrm{X}} \times L_{\mathrm{Y}} \times L_{\mathrm{Z}} = 3\times 1 \times 3$. The Hubbard model has tunneling rates $J_{\mathrm{X}}$=$J_{\mathrm{Y}}$=$J_{\mathrm{Z}}$, interactions $U/J_{\mathrm{Z}}$=$10$, gravity $G/J_{\mathrm{Z}}$=$14$, and SOC phase $\phi$=$ 3\pi/4$. The Rabi frequency is set to values of $\Omega/J_{\mathrm{Z}}$ = $31.5$, $42$ which gives $\mathrm{Z}$ anisotropies of $\Delta_{\mathrm{Z}}$=$-0.5$, $0.5$ respectively via Eq.~\eqref{eq_XXZparams} (left and right panels). Simulations are performed up to times of five units of the anisotropic interaction rate, $t|\chi|$=$5$, with $\chi=-0.6 J_{\mathrm{Z}}$, $-0.2 J_{\mathrm{Z}}$ respectively for the above parameters. Note that the bare Rabi drive $\Omega \hat{S}^{z}$ is manually removed by applying a pulse $e^{+i t\Omega \hat{S}^{z}}$ at the end of the evolution for each time $t$; this can also be done with echo pulses, as discussed in the main text}
\label{fig_HubbardToSpin12}
\end{figure}

We benchmark the validity of the spin-1/2 model by again looking at dynamical evolution of collective spin observables. The same initial state $\ket{\psi_0}$ with all spins pointing along $+x$ on the dressed Bloch sphere is used. Fig.~\ref{fig_HubbardToSpin12} compares the spin-1/2 and Hubbard models using exact numerical time evolution for a small system. The spin-1/2 model captures the Hubbard dynamics well for all studied anisotropies up to timescales of several superexchange times $\chi$ on which significant squeezing can be generated. The parameters used in the simulation are sample values that are within experimental reach. A more specific analysis of candidate platforms and parameters is provided in Appendix~\ref{app_Parameters}.

Note that in our numerical simulations we remove rapid oscillations from the Rabi drive $\Omega \hat{S}^{z}$ by manually applying a final pulse $e^{+i t \Omega \hat{S}^{z}}$ for each time $t$. This drive commutes with the Hamiltonian, but would still need to be accounted for in an experiment. A more sophisticated means of suppressing single-particle terms $\sim\hat{s}_{\vec{r}}^{z}$ from the drive, open boundaries or other uncontrolled shifts is dynamical decoupling via echo pulses.

The simplest form of decoupling is a spin echo, which consists of a single $\pi$ rotation about the $+x$ axis of the Bloch sphere (along which the initial state points) at time $t/2$ halfway through evolution:
\begin{equation}
    \ket{\psi(t)} =e^{-i \hat{H} t/2} e^{-i \hat{\pi} \hat{S}^{x}} e^{-i \hat{H} t/2}\ket{\psi_0}.
\end{equation}
This operation causes single-particle $\hat{s}_{\vec{r}}^{z}$ shifts to undo themselves, up to higher-order shifts from cross-talk between the spins. Since the rotation $\hat{S}^{x}$ is in the dressed basis, it corresponds to a diagonal operator in the original spin basis as discussed in the previous section, and can thus be generated with a detuning $\delta$ of the laser. A detuning corresponds to a Hamiltonian term $\hat{H}_{\mathrm{Detuning}} = \frac{\delta}{2}\sum_{\vec{r}}\left(\hat{n}_{\vec{r},e}-\hat{n}_{\vec{r},g}\right) = \frac{\delta}{2}\sum_{\vec{r}}\left(\hat{a}_{\vec{r},\uparrow}^{\dagger}\hat{a}_{\vec{r},\downarrow} + h.c.\right) = \delta \hat{S}^{x}$. Pulsing a detuning $\delta$ much larger than any other parameter in the system for a time $t \delta = \pi$ efficiently implements an echo pulse.

One can also apply more sophisticated pulse sequences. The simplest example is multiple spin echoes. One can split the evolution time $t$ into $2^{F}$ equal intervals, and apply echo pulses between each interval:
\begin{equation}
\label{eq_multipleEchoes}
    \ket{\psi(t)}= e^{-i \hat{H} t/2^{F}}\left( e^{-i \pi \hat{S}^{x}} e^{-i \hat{H} t/2^F}\right)^{2^F-1}\ket{\psi_0}
\end{equation}

\begin{figure}[h]
\includegraphics[width=0.35\textwidth]{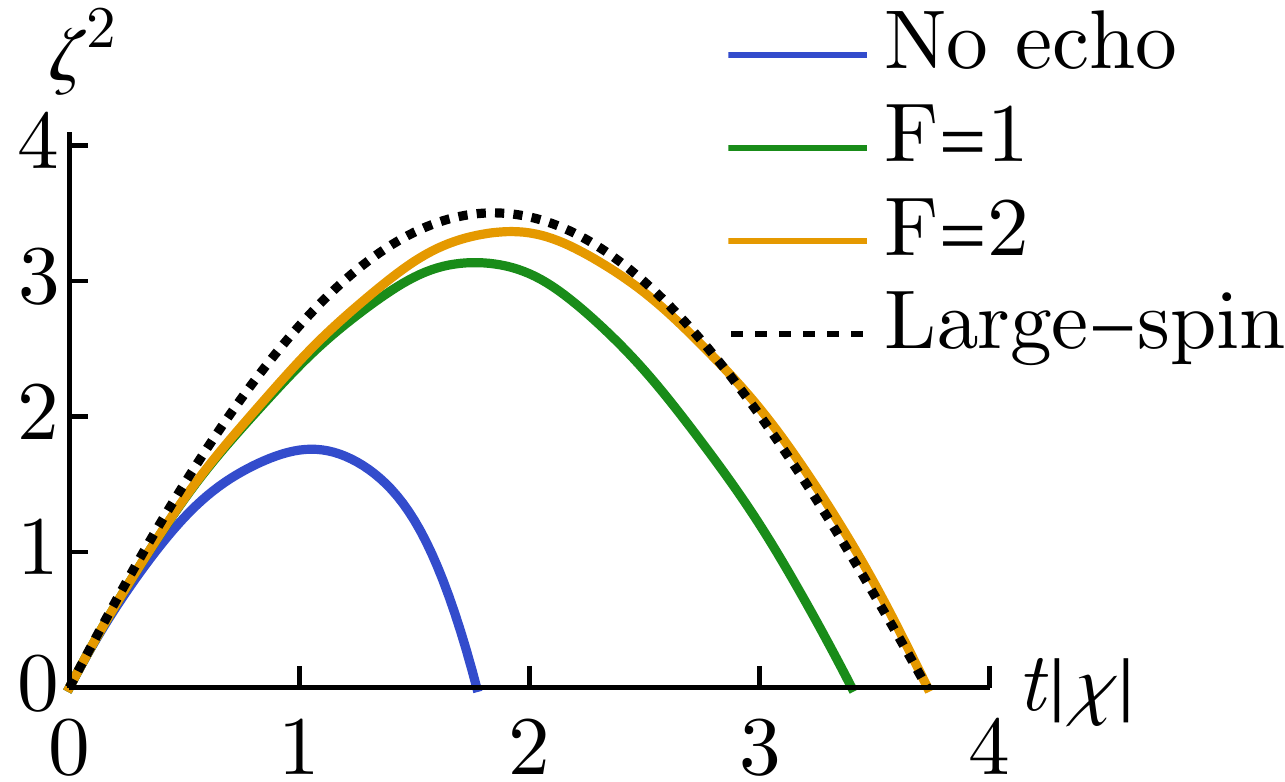}
\centering
\caption{Squeezing generated by the superexchange model $\hat{H}_{\mathrm{Spin}}$ (colored lines) and the large-spin model $\hat{H}_{\mathrm{LS}}$ (black dashed line) using spin echoes to mitigate single-particle shifts, as per Eq.~\eqref{eq_multipleEchoes}. The different colors represent different numbers of spin-echo pulses applied. The system size is $L_{\mathrm{X}}\times L_{\mathrm{Y}}\times L_{\mathrm{Z}}=3\times 1 \times 4$. Parameters are the same as in Fig.~\ref{fig_HubbardToSpin12} for anisotropy $\Delta_{\mathrm{Z}}=-0.5$. The large-spin model uses open boundary conditions along $\mathrm{Z}$ to match the superexchange model, but does not include any single-particle shifts.}
\label{fig_Echo}
\end{figure}

We show that this type of decoupling works in Fig.~\ref{fig_Echo}, plotting the squeezing generated by the spin-1/2 superexchange model $\hat{H}_{\mathrm{Spin}}$ (using open boundaries for which single-particle shifts manifest) for $F=0,1,2$ decoupling sequences, and comparing it to the ideal case of the large-spin model with no single-particle terms. More echo pulses yield higher squeezing that better matches the large-spin result.

\subsection{Time reversal}

Another appealing benefit of our protocol is the ability to not only generate squeezing, but also to reverse it. A common scheme to use spin squeezing for quantum-enhanced metrology is to insert the spin squeezed state as the input of a
Ramsey interferometer, where the system is allowed to accumulate a phase which is detected via spin rotations and population measurements. Performing this last measurement while taking advantage of the  reduced noise variance offered by the spin squeezed  state is challenging since it requires high detection efficiency. This problem can be mitigated by un-squeezing the state after phase accumulation, which amounts to applying the squeezing Hamiltonian with the opposite sign~\cite{davis2016unsqueezing,schulte2020unsqueezing}. Using the time reversal, the noise variance is restored to the one of a coherent spin state, making the final detection simpler. 

In our protocol, the strength of the squeezing Hamiltonian $\chi$ can be flipped by simply quenching the drive frequency from the value $\Omega$ used during squeezing to a new value $\Omega'$. This new value is obtained by solving the equation $\Delta_{\mathrm{Z}}(\Omega) + \Delta_{\mathrm{Z}}(\Omega') = 2$ for $\Omega'$. We have omitted the explicit form of the solution to this equation as it is cumbersome, but a numerical value for a specific set of parameters is straightforward to obtain.

The above quench nevertheless only changes the sign of the anisotropic portion of the Hamiltonian, while keeping the Heisenberg interactions (both along in- and out-of-plane) of the same sign because they are independent of $\Omega$.  We anticipate that the quench will still allow for un-squeezing despite not fully reversing the sign of the Hamiltonian if the system behaves collectively, since in this case the Heisenberg interactions are constants of motion that do not affect the dynamics. In line with our previous results, we expect this requirement to be satisfied for sufficiently large $S$, which is especially appealing for 3D systems where $S$ can be made very large.

We also note that changing the sign of $\chi$ will put one into a parameter regime where the system's $\vec{k} \neq 0$ Bogoliubov modes are unstable because the anisotropy will fall outside the regime $|\Delta_{\mathrm{Z}}| < 1$ (if we were in the stable regime for the squeezing generation). While this would normally lead to non-collective behavior, our prior numerical results show that the strong in-plane interactions favor collective dynamics despite this instability on timescales relevant to spin squeezing.

\subsection{Away from unit filling: role of holes}

The preceding analysis has focused on the ideal case where the initial state $\ket{\psi_0}$ has one atom per site. Real implementations can suffer from holes due to finite entropy conditions. We model the presence of holes by instead considering initial states of the form,
\begin{equation}
    \ket{\psi_0^{\mathrm{doped}}} = \prod_{\vec{r} \notin \mathbbm{H}} \hat{c}_{\vec{r},g}^{\dagger} \ket{0} = \prod_{\vec{r} \notin \mathbbm{H}} \frac{1}{\sqrt{2}}\left(\hat{a}_{\vec{r},\uparrow}^{\dagger} + \hat{a}_{\vec{r},\downarrow}^{\dagger}\right)\ket{0},
\end{equation}
where $\mathbbm{H}$ is a set of sites that have holes. In principle one must take a thermal average over various distributions of $\mathbbm{H}$. In this work we only consider specific sample distributions of randomly-sprinkled holes for simplicity, but do not expect significant differences for collective spin observables provided the system size is large and the filling fraction remains fixed.

The spin-1/2 low-energy description $\hat{H}_{\mathrm{S}}$ of the previous section is no longer valid, as direct motion of atoms into holes must be accounted for with a $t-J$ model~\cite{lee2006tJmodel}. However, the large-spin model can remain valid. Instead of having a spin of size $S$ in every plane, we will have large spins of different lengths $S_{r_{\mathrm{Z}}} = N_{r_{\mathrm{Z}}}/2$, where $N_{r_{\mathrm{Z}}}$ is the number of atoms initially in the plane $r_{\mathrm{Z}}$. The key observation is that the large-spin model $\hat{H}_{\mathrm{LS}}$ has the same equations of motion even for different spin sizes because the spin operators obey the same commutation relations. The main requirement for the model to remain valid is that the angular momentum of each large spin remains fixed at the value it started from during the relevant dynamics.

\begin{figure}[h]
\includegraphics[width=0.48\textwidth]{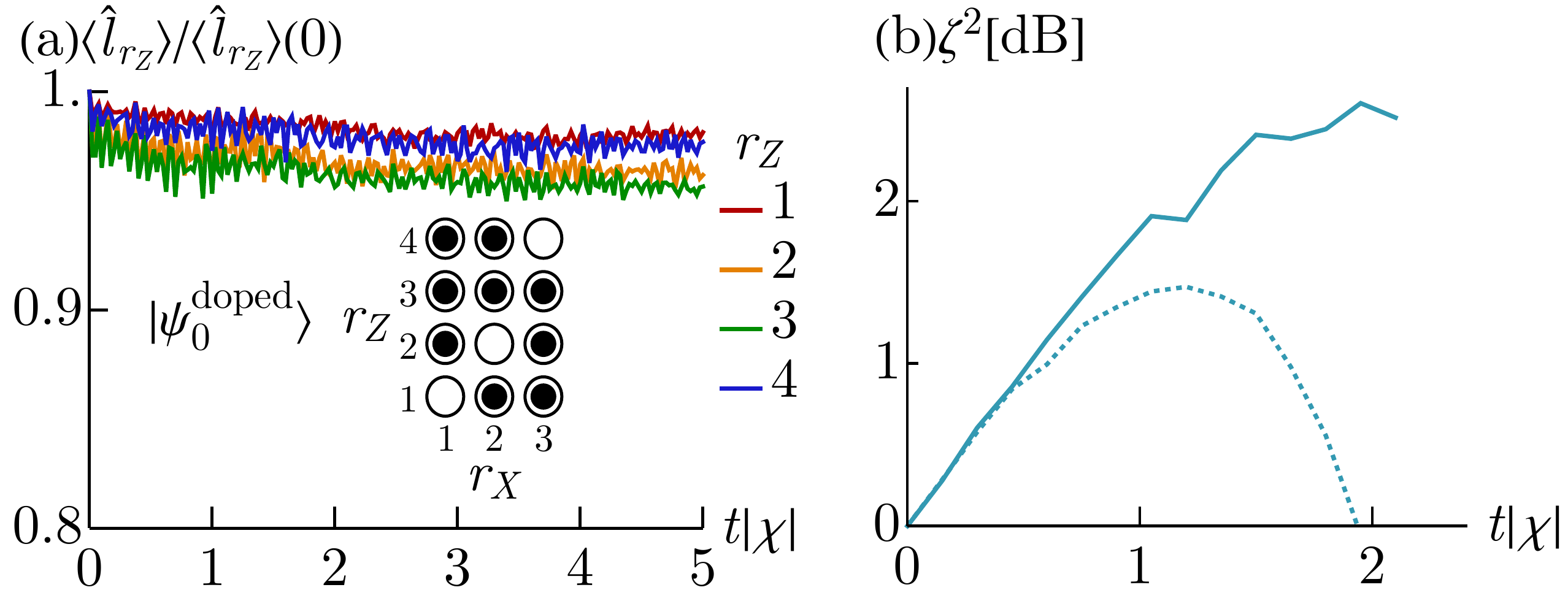}
\centering
\caption{(a) Time evolution of angular momentum $\langle\hat{l}_{r_{\mathrm{Z}}}\rangle$ for each large spin $r_{\mathrm{Z}}$, obtained numerically from the Hubbard model for a system of size $L_{\mathrm{X}}\times L_{\mathrm{Y}}\times L_{\mathrm{Z}} = 3\times 1\times 4$. Each momentum is normalized by its initial value $S_{r_{\mathrm{Z}}}(S_{r_{\mathrm{Z}}}+1)$ with $S_{r_{\mathrm{Z}}} = N_{r_{\mathrm{Z}}}/2$ and $N_{r_{\mathrm{Z}}}$ the number of atoms initially present in each $r_{\mathrm{Z}}$. The initial state $\ket{\psi_0^{\mathrm{doped}}}$ has three holes, as depicted by the empty circles in the inset. The system parameters are otherwise the same as Fig.~\ref{fig_HubbardToSpin12} for anisotropy $\Delta_{\mathrm{Z}}=-0.5$. (b) Spin squeezing generated by the system for the same parameters and initial state as panel (a), reported in decibels $\zeta^2 \text{[dB]} = -10 \log_{10} (\zeta^2)$. An echo pulse is used to mitigate unwanted single-particle terms due to the open boundaries; the dashed line shows the squeezing with no echo pulse.}
\label{fig_Doping}
\end{figure}

The angular momentum of each large spin is given by $\hat{l}_{r_{\mathrm{Z}}}\equiv \vec{S}_{r_{\mathrm{Z}}} \cdot \vec{S}_{r_{\mathrm{Z}}}$, where we write spin operators for simplicity but use the fermionic representations in Eq.~\eqref{eq_SpinOperatorDefinitions} for numerical computations. At time $t=0$ we have $\langle \hat{l}_{r_{\mathrm{Z}}} \rangle (0) = S_{r_{\mathrm{Z}}}(S_{r_{\mathrm{Z}}}+1)$. In Fig.~\ref{fig_Doping}(a) we track the time-evolution of $\langle\hat{l}_{r_{\mathrm{Z}}}\rangle$ for a small system using the Hubbard model for a sample distribution of holes. We find that the angular momentum remains very close to its initial value for each $r_{\mathrm{Z}}$ plane for several superexchange times, indicating that the large-spin mapping still holds.

The physical reason for this validity may also be understood from the spin model picture directly. While atoms can move freely within each $\mathrm{X}-\mathrm{Y}$ plane, the isotropic in-plane interactions will keep the spins aligned (even if the spatial wavefunction becomes complicated). Motion along the $\mathrm{Z}$ direction is suppressed by the linear potential $G$, which creates an energy penalty for atoms to directly hop into vacancies. The lowest-order energy conserving process that can occur along $\mathrm{Z}$ is a virtual double hop, where an atom jumps into the neighbouring plane and then comes back. Such a process occurs with rates of $\sim J_{\mathrm{Z}}^2 \cos^2(\phi/2) /G$ and $\sim J_{\mathrm{Z}}^2 \sin^2(\phi/2) /(\Omega \pm G)$ for dressed-spin-conserving and dressed-spin-flipping hops respectively. These virtual hops manifest as diagonal energy shifts to the respective dressed spin states $\uparrow$, $\downarrow$, yielding non-uniform $\sim \hat{s}_{\vec{r}}^{z}$ spin rotations. If these are smaller than the in-plane Heisenberg interactions, they can also be projected into the Dicke manifold to leading order, yielding global rotations which can be removed via dynamical decoupling.
 
As a more direct benchmark in Fig.~\ref{fig_Doping}(b) we plot the spin squeezing generated by the system for the same doped initial state, solving the Hubbard model directly. This numerical simulation employs an echo pulse ($F=1$ as discussed in Section~\ref{sec_BenchmarkingAndEcho}) to mitigate finite size boundary effects. Despite the very small system size, poor filling fraction, lower 2D dimensionality and strong anisotropy we still find squeezing of several decibels in this "worst-case" test. Larger 3D systems accessible to experiments should be able to generate much higher squeezing.

\section{Conclusions and outlook}
\label{sec_Outlook}

We have shown that multi-dimensional spin models with  a combination of Heisenberg and XXZ models can be used to generate scalable spin squeezing even with  nearest-neighbour couplings. Our approach directly exploits the independent tunability of superexchange interactions along different dimensions of optical lattices, and offers the possibility to perform  time-reversal. The scheme is readily implementable in existing optical lattice clocks used for state-of-the-art metrology, and is robust to common imperfections such as holes. Realization of this protocol has the potential to push existing sensors beyond the current limitations imposed by the standard quantum limit in  non-entangled particles.

It is also interesting to explore experimentally the validity of the large-spin model and its breakdown. There are studies of mixed-dimensional systems~\cite{grusdt2018mixedDimensional} which probe effects such as pairing and string formation. Such large-spin models have also been shown to have connections with canonical models such as the Kitaev chain~\cite{bilitewski2022dipoleLargeSpin}. In our case, properties such as equilibration and propagation of information can change dramatically when passing from large to small spins. Our protocol is quite applicable to such studies, especially given the availability of tools such as quantum gas microscopes~\cite{gross2021microscope} that help probe the microscopic behavior. More generally, understanding the emergence of long-range behavior in effectively short-range systems is crucial to the study of the dynamics of quantum entanglement in many-body physics. The system we describe offers a fantastic opportunity  of performing such studies in a controllable manner.

\textit{Acknowledgements.} This work is supported by the AFOSR grants FA9550-18-1-0319 and FA9550-19-1-0275, by the DARPA and ARO grant W911NF-16-1-0576, by the NSF JILA-PFC PHY-1734006, QLCI-OMA-2016244, by the U.S. Department of Energy, Office of Science, National Quantum Information Science Research Centers Quantum Systems Accelerator, and by NIST.

\bibliographystyle{unsrt}
\bibliography{BiblioSpin}

\clearpage
\onecolumngrid
\appendix

\section{Holstein-Primakoff approximation}
\label{app_LargeSpin}
\subsection{Quadratic theory}

This Appendix explores the validity of the mapping from the 3D spin-1/2 model $\hat{H}_{\mathrm{S}}$ to the large-spin model $\hat{H}_{\mathrm{LS}}$ using a Holstein-Primakoff approach. We start with the spin-1/2 model, assuming periodic boundaries, equal and isotropic in-plane couplings $V_{\mathrm{X}} = V_{\mathrm{Y}} = V_{\mathrm{XY}}$ and $\Delta_{\mathrm{X}} = \Delta_{\mathrm{Y}} = 1$,
\begin{equation}
\hat{H}_{\mathrm{S}} = V_{\mathrm{XY}} \sum_{\vec{r}} \left(\vec{s}_{\vec{r}} \cdot \vec{s}_{\vec{r}+\vec{\mathrm{X}}} +  \vec{s}_{\vec{r}} \cdot \vec{s}_{\vec{r}+\vec{\mathrm{Y}}}\right) + V_{\mathrm{Z}}\sum_{\vec{r}}\left[\hat{s}_{\vec{r}}^{x} \hat{s}_{\vec{r} + \vec{\mathrm{Z}}}^{x}+\hat{s}_{\vec{r}}^{y} \hat{s}_{\vec{r} + \vec{\mathrm{Z}}}^{y} + \Delta_{\mathrm{Z}} \hat{s}_{\vec{r}}^{z}\hat{s}_{\vec{r}+\vec{\mathrm{Z}}}^{z}\right],
\end{equation}
All couplings are made positive $V_{\mathrm{XY}}, V_{\mathrm{Z}} > 0$. For simplicity we also set the system size equal along all directions, $L_{\mathrm{X}}=L_{\mathrm{Y}} = L_{\mathrm{Z}}\equiv L$, with a total of $N = L^3$ spins.

We start with a lowest-order Holstein-Primakoff transformation upon the spin-1/2 operators:
\begin{equation}
\begin{aligned}
    \hat{s}_{\vec{r}}^{x} &= \frac{1}{2} - \hat{b}_{\vec{r}}^{\dagger}\hat{b}_{\vec{r}},\\
    \hat{s}_{\vec{r}}^{y} &= - \frac{i}{2 }\left(\hat{b}_{\vec{r}} - \hat{b}_{\vec{r}}^{\dagger}\right),\\
    \hat{s}_{\vec{r}}^{z} &= - \frac{1}{2}\left(\hat{b}_{\vec{r}} + \hat{b}_{\vec{r}}^{\dagger}\right),
\end{aligned}
\end{equation}
where $\hat{b}_{\vec{r}}$ are bosonic modes satisfying commutation relations of $[\hat{b}_{\vec{r}},\hat{b}_{\vec{r}'}^{\dagger}] = \delta_{\vec{r},\vec{r}'}$. These definitions are chosen to match the initial state $\ket{\psi_0}$ with each spin pointing along the $+x$ direction of the Bloch sphere, such that at $t=0$ all bosonic modes are in vacuum $\ket{0}$. While this transformation is only formally exact in the limit of spin size going to infinity, it should still give a valid description of the system dynamics provided that the number of excitations is sufficiently small.

We start with the quadratic non-interacting theory. Writing $\hat{H}_{\mathrm{S}}$ in terms of the bosonic operators and keeping only terms of quadratic order yields:
\begin{equation}
\begin{aligned}
    \hat{H}_{\mathrm{S}}^{\mathrm{non-int}} = V_{\mathrm{XY}}&\sum_{\vec{r}}\left[-2\hat{n}_{\vec{r}} + \frac{1}{2}\left(\hat{b}_{\vec{r}}^{\dagger}\hat{b}_{\vec{r}+\vec{\mathrm{X}}} + h.c.\right) + \frac{1}{2}\left(\hat{b}_{\vec{r}}^{\dagger}\hat{b}_{\vec{r}+\vec{\mathrm{Y}}} + h.c.\right)\right] \\
    + V_{\mathrm{Z}}&\sum_{\vec{r}}\left[-\hat{n}_{\vec{r}} + \frac{\Delta_{\mathrm{Z}}+1}{4}\left(\hat{b}_{\vec{r}}^{\dagger}\hat{b}_{\vec{r}+\vec{\mathrm{Z}}} + h.c.\right) + \frac{\Delta_{\mathrm{Z}}-1}{4}\left(\hat{b}_{\vec{r}}\hat{b}_{\vec{r}+\vec{\mathrm{Z}}}+h.c.\right)\right].
\end{aligned}
\end{equation}
Making a Fourier transformation $\hat{b}_{\vec{r}} = \frac{1}{\sqrt{N}}\sum_{\vec{k}}e^{i \vec{k} \cdot \vec{r}} \hat{b}_{\vec{k}}$ with $\vec{k}=(k_{\mathrm{X}},k_{\mathrm{Y}},k_{\mathrm{Z}})$ gives
\begin{equation}
\begin{aligned}
    \hat{H}_{\mathrm{S}}^{\mathrm{non-int}} &= \sum_{\vec{k}}\epsilon_{\vec{k}}\hat{n}_{\vec{k}} + \sum_{\vec{k}}\frac{\Delta_{\vec{k}}}{2}\left(\hat{b}_{\vec{k}}\hat{b}_{-\vec{k}}+h.c.\right),
\end{aligned}
\end{equation}
with coefficients of,
\begin{equation}
\label{eq_3DHolsteinKCoefficients}
\begin{aligned}
\epsilon_{\vec{k}} &= -2V_{\mathrm{XY}} \left[\sin^2 \left(\frac{k_{\mathrm{X}}}{2}\right) + \sin^2 \left(\frac{k_{\mathrm{Y}}}{2}\right)\right]-2 V_{\mathrm{Z}}\sin^2 \left(\frac{k_{\mathrm{Z}}}{2}\right) + V_{\mathrm{Z}} \frac{\Delta_{\mathrm{Z}}-1}{2}\cos(k_{\mathrm{Z}}),\\
\Delta_{\vec{k}} &= V_{\mathrm{Z}}\frac{\Delta_{\mathrm{Z}}-1}{2}\cos(k_{\mathrm{Z}}).
\end{aligned}
\end{equation}
This quadratic system can be solved analytically with a Bogoliubov transformation. The resulting eigenmodes have energies $\sqrt{\epsilon_{\vec{k}}^2 - \Delta_{\vec{k}}^2}$. The time-dependent excitation number of a given $\vec{k}$ mode is,
\begin{equation}
\label{eq_BogoliubovPopulation}
\langle\hat{b}_{\vec{k}}^{\dagger}\hat{b}_{\vec{k}}\rangle = \frac{\Delta_{\vec{k}}^2}{\epsilon_{\vec{k}}^2 - \Delta_{\vec{k}}^2} \sin^2 \left(t \sqrt{\epsilon_{\vec{k}}^2 - \Delta_{\vec{k}}^2}\right).
\end{equation}
For a parametrically stable mode with $|\epsilon_{\vec{k}}| > |\Delta_{\vec{k}}|$ the population will oscillate sinusoidally, whereas an unstable mode with $|\epsilon_{\vec{k}}| < |\Delta_{\vec{k}}|$ will undergo exponential growth. The borderline case $|\epsilon_{\vec{k}}| = |\Delta_{\vec{k}}|$ tends to yield polynomial growth of population with time. In our case, using the explicit forms of $\epsilon_{\vec{k}}$, $\Delta_{\vec{k}}$ one can show that for anisotropy $|\Delta_{\mathrm{Z}}|<1$, all modes with $\vec{k} \neq 0$ are parametrically stable, whereas the converse regime $|\Delta_{\mathrm{Z}}| > 1$ leads to some unstable modes. The $\vec{k}=0$ mode always sits at the edge of instability with $\epsilon_{0} = \Delta_{0}$ regardless of the anisotropy.

The large-spin model requires a lack of excitations with non-zero in-plane quasimomentum $(k_{\mathrm{X}}, k_{\mathrm{Y}}) \neq 0$, which ensures that each plane remains collective during the dynamics. In what follows, we will derive an analytic bound upon the total number of such excitations. This bound will be used to determine when the large-spin model mapping is valid, and when scalable squeezing can be generated.

\subsection{Bound on spin-wave excitations in the stable regime}

The total number of excitations is,
\begin{equation}
N_{\mathrm{exc}}=\sum_{\vec{k}} \langle\hat{b}_{\vec{k}}^{\dagger}\hat{b}_{\vec{k}}\rangle.
\end{equation}
The relevant quantity for the validity of the large-spin model is a restricted sum over modes with $(k_{\mathrm{X}}, k_{\mathrm{Y}}) \neq 0$. However, we will show that the above sum can be bounded when summing over all $\vec{k} \neq 0$ modes. 
If all such modes are parametrically stable, $|\epsilon_{\vec{k}}| > |\Delta_{\vec{k}}|$, requiring $|\Delta_{\mathrm{Z}}|<1$, we can bound the time-dependent sinusoidal term in the population,
\begin{equation}
N_{\mathrm{exc}}=\sum_{\vec{k}} \frac{\Delta_{\vec{k}}^2}{\epsilon_{\vec{k}}^2 - \Delta_{\vec{k}}^2} \sin^2 \left(t \sqrt{\epsilon_{\vec{k}}^2 - \Delta_{\vec{k}}^2}\right) \leq \sum_{\vec{k}}\frac{\Delta_{\vec{k}}^2}{\epsilon_{\vec{k}}^2 - \Delta_{\vec{k}}^2}.
\end{equation}
In writing the above expression we have neglected the time-dependent growth of the $\vec{k} = 0$ mode. This growth drives squeezing generation, and does not affect the validity of the large-spin model for our purposes; we will explicitly solve the dynamics of the $\vec{k}=0$ mode further on.

We insert the explicit forms of the coefficients $\epsilon_{\vec{k}}$ and $\Delta_{\vec{k}}$ and simplify the summand,
\small
\begin{equation}
N_{\mathrm{exc}}\leq  \frac{\lambda^2(\Delta_{\mathrm{Z}}-1)^2}{16 }\sum_{\vec{k}}\frac{\cos^2 k_{\mathrm{Z}}}{\left[\sin^2\left(\frac{k_{\mathrm{X}}}{2}\right)+\sin^2\left(\frac{k_{\mathrm{Y}}}{2}\right)+\frac{\lambda}{2}(1-\cos k_{\mathrm{Z}})\right]\left[\sin^2\left(\frac{k_{\mathrm{X}}}{2}\right)+\sin^2\left(\frac{k_{\mathrm{Y}}}{2}\right)+\frac{\lambda}{2}(1-\Delta_{\mathrm{Z}}\cos k_{\mathrm{Z}})\right]},
\end{equation}
\normalsize
where we have defined,
\begin{equation}
    \lambda = V_{\mathrm{Z}}/V_{\mathrm{XY}}.
\end{equation}
In the limit of $L \gg 1$ we re-write the Fourier sum as an integral $\sum_{\vec{k}} \to \frac{L^3}{(2\pi)^3}\int d^{3}\vec{k}$,
\small
\begin{equation}
\label{eq_FullSumInStableRegimeInt}
N_{\text{exc}}\leq L^3\frac{\lambda^2(\Delta_{\mathrm{Z}}-1)^2}{128\pi^3}\int d^3\vec{k} \frac{\cos^2 k_{\mathrm{Z}}}{\left[\sin^2\left(\frac{k_{\mathrm{X}}}{2}\right)+\sin^2\left(\frac{k_{\mathrm{Y}}}{2}\right)+\frac{\lambda}{2}(1-\cos k_{\mathrm{Z}})\right]\left[\sin^2\left(\frac{k_{\mathrm{X}}}{2}\right)+\sin^2\left(\frac{k_{\mathrm{Y}}}{2}\right)+\frac{\lambda}{2}(1-\Delta_{\mathrm{Z}}\cos k_{\mathrm{Z}})\right]}.
\end{equation}
\normalsize
We seek to evaluate the integral,
\begin{equation}
\begin{aligned}
I \equiv \int d^3\vec{k} \frac{\cos^2 k_{\mathrm{Z}}}{\left[\sin^2\left(\frac{k_{\mathrm{X}}}{2}\right)+\sin^2\left(\frac{k_{\mathrm{Y}}}{2}\right)+\frac{\lambda}{2}(1-\cos k_{\mathrm{Z}})\right]\left[\sin^2\left(\frac{k_{\mathrm{X}}}{2}\right)+\sin^2\left(\frac{k_{\mathrm{Y}}}{2}\right)+\frac{\lambda}{2}(1-\Delta_{\mathrm{Z}}\cos k_{\mathrm{Z}})\right]}.
\end{aligned}
\end{equation}
The integration over $k_{\mathrm{X}}$, $k_{\mathrm{Y}}$ can be done analytically, yielding,
\begin{equation}
I=\frac{64\pi}{\lambda(1-\Delta_{\mathrm{Z}})} \int_0^{\pi} d k_{\mathrm{Z}}\cos k_{\mathrm{Z}}\left(\frac{K\left\{\frac{-4}{\lambda \left(1-\cos k_{\mathrm{Z}}\right)\left[4+\lambda (1-\cos k_{\mathrm{Z}})\right]}\right\}}{\sqrt{1+\frac{\lambda}{2}(1-\cos k_{\mathrm{Z}})}}-\frac{K\left\{\frac{-4}{\lambda\left(1-\Delta_{\mathrm{Z}}\cos k_{\mathrm{Z}}\right)\left[4+\lambda (1-\Delta_{\mathrm{Z}}\cos k_{\mathrm{Z}})\right]}\right\}}{\sqrt{1+\frac{\lambda}{2}(1-\Delta_{\mathrm{Z}}\cos k_{\mathrm{Z}})}}\right),
\end{equation}
where $K\{x\}$ is the elliptic integral of the first kind. This remaining integral cannot be solved exactly. However, we seek to work in the parameter regime where the in-plane coupling is much stronger than the out-of-plane one,
\begin{equation}
    \lambda \ll 1.
\end{equation}
In this regime the arguments of the elliptic integral functions are very large and negative for all $k_{\mathrm{Z}}$. We can employ the known Taylor expansion of the elliptic integral about $-\infty$,
\begin{equation}
    K\{-x\} = \frac{4\log 2 + \log x}{\sqrt{x}} + \mathcal{O}\left(\frac{1}{x^{3/2}}\right),\>\>\>\>x > 0.
\end{equation}
Applying this expansion to the elliptic integrals and simplifying the integrand yields,
\begin{equation}
    I =\frac{16\pi}{\lambda(1-\Delta_{\mathrm{Z}})}\int_0^{\pi} d k_{\mathrm{Z}}\cos k_{\mathrm{Z}}\log \left[\frac{(1-\Delta_{\mathrm{Z}}\cos k_{\mathrm{Z}})\left[4+\lambda \left(1-\Delta_{\mathrm{Z}}\cos k_{\mathrm{Z}}\right)\right]}{(1-\cos k_{\mathrm{Z}})\left[4+\lambda \left(1-\cos k_{\mathrm{Z}}\right)\right]}\right] + \mathcal{O}\left(\lambda^{3/2}\right).
\end{equation}
We next Taylor expand the integrand to zeroth order in $\lambda$, yielding,
\begin{equation}
    I =\frac{16\pi}{\lambda(1-\Delta_{\mathrm{Z}})}\int_0^{\pi} d k_{\mathrm{Z}}\cos k_{\mathrm{Z}}\log \left(\frac{1-\Delta_{\mathrm{Z}}\cos k_{\mathrm{Z}}}{1-\cos k_{\mathrm{Z}}}\right) + \mathcal{O}\left(\lambda^{0}\right).
\end{equation}
This integral may now be evaluated analytically,
\begin{equation}
    I =\frac{16\pi^2}{\lambda(1-\Delta_{\mathrm{Z}})}\left(1 - \frac{1-\sqrt{1-\Delta_{\mathrm{Z}}^2}}{\Delta_{\mathrm{Z}}}\right) + \mathcal{O}(\lambda^0).
\end{equation}
Inserting the value of the integral into Eq.~\eqref{eq_FullSumInStableRegimeInt} gives us a bound on the excitation number,
\begin{equation}
\begin{aligned}
\label{eq_Bound}
    N_{\mathrm{exc}} &\leq \frac{\lambda (1-\Delta_{\mathrm{Z}}) L^3}{8\pi}\left(1-\frac{1-\sqrt{1-\Delta_{\mathrm{Z}}^2}}{\Delta_{\mathrm{Z}}}\right) + \mathcal{O}(\lambda^2)\\
    &\sim \eta N,
\end{aligned}
\end{equation}
where on the second line we have used the modified small parameter $\eta = \lambda (1- \Delta_{\mathrm{Z}}) = |V_{\mathrm{Z}}(\Delta_{\mathrm{Z}}-1)/V_{\mathrm{XY}}|$ that was employed in the main text. The anisotropy-dependent factor inside the large brackets does not exhibit any divergences or singular behavior in the regime of interest.

If $\eta \ll 1$, the contribution from stable Bogoliubov modes will not significantly contribute to the expectation values of collective spin observables $\hat{S}^{\alpha}$, $\hat{S}^{\alpha}\hat{S}^{\beta}$ with $\hat{S}^{\alpha} = \sum_{\vec{r}}\hat{s}_{\vec{r}}^{\alpha}$, because these observables are extensive and also scale with $N$. This argument can break down for 
specific combinations of observables where cancellation between extensive quantities can occur, and for contributions from the $\vec{k}=0$ mode (because its time dependence is not bounded). However, we show in the next section that such contributions do not inhibit squeezing generation.

\subsection{Zero mode}

The $\vec{k}=0$ mode sits on the edge of parametric instability and generates squeezing dynamics.
The quadratic Holstein-Primakoff Hamiltonian for this mode is,
\begin{equation}
\hat{H}_{\mathrm{S}}^{(\vec{k} = 0)}=\frac{\chi}{2} \left(\hat{b}_{0}^{\dagger}\hat{b}_{0} + \frac{1}{2} \hat{b}_{0}\hat{b}_{0} + \frac{1}{2}\hat{b}_{0}^{\dagger}\hat{b}_{0}^{\dagger}\right).
\end{equation}

The observables relevant to spin squeezing are collective spin operators. The specific necessary operators (as we will show further on) are $\hat{S}^{x}$, $\hat{S}^{y}\hat{S}^{y}$ and  $\text{Re}[\hat{S}^{y}\hat{S}^{z}]=(\hat{S}^{y}\hat{S}^{z}+\hat{S}^{z}\hat{S}^{y})/2$. These operators may be written in the Holstein-Primakoff picture,
\begin{equation}
\begin{aligned}
\hat{S}^{x} &= \frac{N}{2} - \sum_{\vec{k}} \hat{b}_{\vec{k}}^{\dagger}\hat{b}_{\vec{k}} \approx \frac{N}{2} - \hat{b}_{0}^{\dagger}\hat{b}_{0},\\
\hat{S}^{y}\hat{S}^{y} &=\frac{N}{4}\left(1 + 2 \hat{b}_{0}^{\dagger}\hat{b}_{0}- \hat{b}_{0}\hat{b}_{0} - \hat{b}_{0}^{\dagger}\hat{b}_{0}^{\dagger}\right),\\
\text{Re}[\hat{S}^{y}\hat{S}^{z}] &= \frac{i N}{4} \left(\hat{b}_{0}\hat{b}_{0}-\hat{b}_{0}^{\dagger}\hat{b}_{0}^{\dagger}\right).
\end{aligned}
\end{equation}
The two-point operators involve only the $\vec{k}=0$ mode inherently.  The one-point operator $\hat{S}^{x}$ also contains contributions from $\vec{k} \neq 0$, but if the system is in the stable regime $|\Delta_{\mathrm{Z}}|< 1$ and the interaction ratio satisfies $\eta \ll 1$ these contributions can be neglected in the large system size limit $L \gg 1$.

The Heisenberg equations of motion for the relevant bosonic operators may be directly computed from $\hat{H}_{\mathrm{S}}^{(\vec{k}=0)}$,
\begin{equation}
\frac{d}{dt}\left(\begin{array}{c} \langle\hat{b}_{0}^{\dagger}\hat{b}_{0}\rangle\\ \langle\hat{b}_{0}\hat{b}_{0}\rangle\\ \langle\hat{b}_{0}^{\dagger}\hat{b}_{0}^{\dagger}\rangle \end{array}\right) =i \frac{\chi}{2}\left(\begin{array}{ccc} 0 & 1 & -1 \\ -2 & -2 & 0 \\ 2 & 0 & 2  \end{array}\right)\left(\begin{array}{c} \langle\hat{b}_{0}^{\dagger}\hat{b}_{0}\rangle\\ \langle\hat{b}_{0}\hat{b}_{0}\rangle\\ \langle\hat{b}_{0}^{\dagger}\hat{b}_{0}^{\dagger}\rangle\end{array}\right) +\frac{i \chi}{2}\left(\begin{array}{c}0 \\ -1 \\ 1 \end{array}\right).
\end{equation}
Using the initial conditions of no excitations $\langle\hat{b}_{0}^{\dagger}\hat{b}_{0}\rangle(0) = \langle\hat{b}_{0}\hat{b}_{0}\rangle(0) = \langle\hat{b}_{0}^{\dagger}\hat{b}_{0}^{\dagger}\rangle(0)=0$, these equations yield explicit solutions:
\begin{equation}
\begin{aligned}
\label{eq_K0ModeExpectationValues}
\langle\hat{b}_{0}^{\dagger}\hat{b}_{0}\rangle &=\frac{1}{4} (\chi t)^2,\\
\langle\hat{b}_{0}\hat{b}_{0}\rangle &=-\frac{1}{4}(\chi t)^2 - \frac{i}{2}\chi t,\\
\langle\hat{b}_{0}^{\dagger}\hat{b}_{0}^{\dagger}\rangle &=-\frac{1}{4}(\chi t)^2 + \frac{i} {2}\chi t.
\end{aligned}
\end{equation}
The relevant collective spin expectation values then read:
\begin{equation}
\begin{aligned}
\langle \hat{S}^{x}\rangle &= \frac{N}{2} - \frac{1}{4}(\chi t)^2,\\
\langle\hat{S}^{y} \hat{S}^{y}\rangle &= \frac{N}{4}\left[1+(\chi t)^2\right],\\
\text{Re}[\langle\hat{S}^{y} \hat{S}^{z}\rangle]&= \frac{N}{4}\chi t.
\end{aligned}
\end{equation}

With these expressions in hand we compute the squeezing. For the initial state $\ket{\psi_0}$, assuming periodic boundary conditions the squeezing can be written as,
\begin{equation}
\begin{aligned}
\zeta^2 &= N\text{min}_{\theta} \frac{\langle \Delta \vec{S}_{\perp \theta}\rangle}{|\langle\vec{S}\rangle|^2}\\
&=\frac{N}{\langle \hat{S}^{x}\rangle^2} \text{min}_{\theta} \bigg[\frac{N}{4} \sin^2(\theta) + \langle\hat{S}^{y}\hat{S}^{y}\rangle \cos^2(\theta) -\sin(2\theta)\text{Re}[\langle\hat{S}^{y} \hat{S}^{z}\rangle]\bigg].
\end{aligned}
\end{equation}
Inserting the solutions from earlier and minimizing the expression yields,
\begin{equation}
    \zeta^2 = \frac{N^2 \left(1-|\chi| t \left[\sqrt{1+ \frac{1}{4}(\chi t)^2}-\frac{1}{2}|\chi| t\right]\right)}{\left[N-\frac{1}{2}(\chi t)^2\right]^2}.
\end{equation}
In the limit $N \to \infty$ this expression simplifies to,
\begin{equation}
    \zeta^2 = \frac{1}{\left(\sqrt{1+\frac{\chi^2 t^2}{4}}+ \frac{|\chi| t}{2}\right)^2} + \mathcal{O}\left(\frac{1}{N}\right),
\end{equation}
as in the main text.

We can also explicitly show that corrections from $\vec{k} \neq 0$ modes to $\hat{S}^{x}$ do not change the squeezing scaling. Indeed, if  these corrections are bounded by $\sim \eta N$ as shown previously we can make the replacement of,
\begin{equation}
    \langle\hat{S}^{x}\rangle \to \langle\hat{S}^{x}\rangle - \eta N,
\end{equation}
the squeezing becomes,
\begin{equation}
\begin{aligned}
\zeta^2 &= \frac{N}{\left(\langle \hat{S}^{x}\rangle-\eta N\right)^2} \text{min}_{\theta} \bigg[\frac{N}{4} \sin^2(\theta) + \langle\hat{S}^{y}\hat{S}^{y}\rangle \cos^2(\theta) -\sin(2\theta)\text{Re}[\langle\hat{S}^{y} \hat{S}^{z}\rangle]\bigg]\\
&= \left(\frac{N}{\langle \hat{S}^{x}\rangle^2} +2 \eta \frac{N^2}{\langle \hat{S}^{x}\rangle^3} +\mathcal{O}(\eta^2)\right)\text{min}_{\theta} \bigg[\frac{N}{4} \sin^2(\theta) + \langle\hat{S}^{y}\hat{S}^{y}\rangle \cos^2(\theta) -\sin(2\theta)\text{Re}[\langle\hat{S}^{y} \hat{S}^{z}\rangle]\bigg]
\end{aligned}
\end{equation}
The leading-order correction scales as $\sim \eta N^2 / \langle\hat{S}^{x}\rangle^3$. For one-axis twisting, at the optimal squeezing time the system still has non-vanishing macroscopic contrast $\langle\hat{S}^{x}\rangle\sim N$. The corrections thus scale as a $\sim \eta/N$ adjustment to the prefactor, which will not change the resulting scaling of the optimal squeezing as $\zeta_{\mathrm{min}}\sim N^{-2/3}$.

\section{Superexchange model derivation}
\label{app_Superexchange}

\subsection{Fermionic model}

In this Appendix we derive the spin-1/2 superexchange model for fermionic atoms provided in the main text. We start by writing the Hubbard model $\hat{H}$ in the dressed basis for two atoms populating two neighbouring lattice sites $r_{\mathrm{Z}}\in \{1,2\}$ along the $\mathrm{Z}$ direction. It is convenient to express this model in a basis of Fock states $\{\ket{n_{1,\uparrow},n_{1,\downarrow},n_{2,\uparrow},n_{2,\downarrow}}\}$, with $n_{r_{\mathrm{Z}},\tilde{\sigma}}\in \{0,1\}$ the number of atoms on site $r_{\mathrm{Z}}$ with dressed spin $\tilde{\sigma}\in \{\uparrow,\downarrow\}$. There are a total of 6 Fock states given by $\{\ket{1,1,0,0},\ket{1,0,1,0},\ket{1,0,0,1},\ket{0,1,1,0},\ket{0,1,0,1},\ket{0,0,1,1}\}$. The two-site Hubbard Hamiltonian in this basis can be written as,
\begin{equation}
    \hat{H}^{(2)} = \hat{H}_{\mathrm{Diagonal}}^{(2)} + \hat{H}_{\mathrm{Tunneling}}^{(2)}.
\end{equation}
The first term $\hat{H}_{\mathrm{Diagonal}}^{(2)} = \hat{H}_{\mathrm{Interaction}}^{(2)} + \hat{H}_{\mathrm{Gravity}}^{(2)} + \hat{H}_{\mathrm{Drive}}^{(2)}$ contains the on-site Hubbard interactions, gravity, and laser drive, all of which are diagonal in the dressed basis,
\begin{equation}
    \hat{H}_{\mathrm{Diagonal}}^{(2)} = \hat{H}_{\mathrm{Interaction}}^{(2)} + \hat{H}_{\mathrm{Gravity}}^{(2)} + \hat{H}_{\mathrm{Drive}}^{(2)}=  \left(
\begin{array}{cccccc}
 U+2G & 0 & 0 & 0 & 0 & 0 \\
 0 & \Omega+3 G  & 0 & 0 & 0 & 0 \\
 0 & 0 & 3 G & 0 & 0 & 0 \\
 0 & 0 & 0 & 3 G & 0 & 0 \\
 0 & 0 & 0 & 0 & -\Omega+3 G  & 0 \\
 0 & 0 & 0 & 0 & 0 & U+4 G \\
\end{array}
\right).
\end{equation}
The other term is the tunneling,
\begin{equation}
    \hat{H}_{\mathrm{Tunneling}}^{(2)} = \left(
\begin{array}{cccccc}
0 & i J_{\mathrm{Z}} \sin \left(\frac{\phi }{2}\right) & -J_{\mathrm{Z}} \cos \left(\frac{\phi }{2}\right) & J_{\mathrm{Z}} \cos \left(\frac{\phi }{2}\right) & -i J_{\mathrm{Z}} \sin
   \left(\frac{\phi }{2}\right) & 0 \\
 -i J_{\mathrm{Z}} \sin \left(\frac{\phi }{2}\right) & 0 & 0 & 0 & 0 & -i J_{\mathrm{Z}} \sin \left(\frac{\phi }{2}\right) \\
 -J_{\mathrm{Z}} \cos \left(\frac{\phi }{2}\right) & 0 & 0 & 0 & 0 & -J_{\mathrm{Z}} \cos \left(\frac{\phi }{2}\right) \\
 J_{\mathrm{Z}} \cos \left(\frac{\phi }{2}\right) & 0 & 0 & 0 & 0 & J_{\mathrm{Z}} \cos \left(\frac{\phi }{2}\right) \\
 i J_{\mathrm{Z}} \sin \left(\frac{\phi }{2}\right) & 0 & 0 & 0 & 0 & i J_{\mathrm{Z}} \sin \left(\frac{\phi }{2}\right) \\
 0 & i J_{\mathrm{Z}} \sin \left(\frac{\phi }{2}\right) & -J_{\mathrm{Z}} \cos \left(\frac{\phi }{2}\right) & J_{\mathrm{Z}} \cos \left(\frac{\phi }{2}\right) & -i J_{\mathrm{Z}} \sin
   \left(\frac{\phi }{2}\right) & 0 \\
\end{array}
\right).
\end{equation}

The spin model is obtained by treating $\hat{H}_{\mathrm{Tunneling}}^{(2)}$ as a perturbation, applying standard second-order Schrieffer-Wolff perturbation theory to trace out the doubly occupied states $\ket{1,1,0,0}$, $\ket{0,0,1,1}$, and then mapping the remaining effective Hamiltonian to spin-1/2 degrees of freedom. The Schrieffer-Wolff generator is defined as,
\begin{equation}
    \hat{S} = \sum_{n,n'} \frac{\bra{n}\hat{H}_{\mathrm{Tunneling}}^{(2)}\ket{n'}}{\bra{n}\hat{H}_{\mathrm{Diagonal}}^{(2)}\ket{n}- \bra{n'}\hat{H}_{\mathrm{Diagonal}}^{(2)}\ket{n'}} \ket{n}\bra{n'},
\end{equation}
where $n,n'$ each run over the 6 basis states. We also define the projector $\hat{P}$ onto the non-doubly-occupied states,
\begin{equation}
\hat{P} = \left(\begin{array}{cccccc}
0 & 0 & 0 & 0 & 0 & 0\\
0 & 1 & 0 & 0 & 0 & 0\\
0 & 0 & 1 & 0 & 0 & 0\\
0 & 0 & 0 & 1 & 0 & 0\\
0 & 0 & 0 & 0 & 1 & 0\\
0 & 0 & 0 & 0 & 0 & 0\\
\end{array}\right).
\end{equation}
The second-order perturbative Hamiltonian can then be written as,
\begin{equation}
    \hat{H}_{\mathrm{Spin}}^{(2)} = \hat{P} \hat{H}_{\mathrm{Diagonal}}^{(2)} \hat{P} + \frac{1}{2}\hat{P}[\hat{S},\hat{H}_{\mathrm{Tunneling}}^{(2)}]\hat{P}.
\end{equation}
The $\ket{1,1,0,0}$ and $\ket{0,0,1,1}$ states (the first and last basis elements) are dropped, leaving a $4 \times 4$ matrix.

Before writing the resulting matrix, we observe that the diagonal energies from $\hat{H}_{\mathrm{Diagonal}}^{(2)}$ (given by $3G+\Omega, 3G, 3G, 3G-\Omega$) are much larger than any matrix elements obtained from the perturbation. We thus drop any off-diagonal matrix elements of $\hat{H}_{\mathrm{Spin}}^{(2)}$ that couple states with different diagonal energies. The resulting Hamiltonian reads,
\begin{equation}
    \hat{H}_{\mathrm{Spin}}^{(2)} =\left(
\begin{array}{cccc}
 \Omega-\frac{2J_{\mathrm{Z}}^2 (U-\Omega) \sin^2\left(\frac{\phi}{2}\right)}{(U-\Omega)^2-G^2} & \cdot & \cdot & \cdot \\
 \cdot & -\frac{2J_{\mathrm{Z}}^2 U \cos^2\left(\frac{\phi}{2}\right)}{U^2-G^2} & \frac{2J_{\mathrm{Z}}^2 U \cos^2\left(\frac{\phi}{2}\right)}{U^2-G^2}  & \cdot \\
 \cdot & \frac{2J_{\mathrm{Z}}^2 U \cos^2 \left(\frac{\phi}{2}\right)}{U^2-G^2} & -\frac{2J_{\mathrm{Z}}^2 U \cos^2\left(\frac{\phi}{2}\right)}{U^2-G^2} & \cdot \\
 \cdot & \cdot & \cdot & -\Omega-\frac{2J_{\mathrm{Z}}^2 (U+\Omega) \sin^2 \left(\frac{\phi}{2}\right)}{(U+\Omega)^2-G^2} \\
\end{array}
\right) + \text{(const)},
\end{equation}
where $\cdot$ denotes terms that were dropped, and (const) refers to terms proportional to the identity.

This Hamiltonian can be expressed in terms of spin-1/2 operators $\hat{s}_{r_{\mathrm{Z}}}^{\alpha}$ for the two sites $r_{\mathrm{Z}}=1,2$,
\begin{equation}
\begin{aligned}
\hat{s}_{1}^{\alpha} &= \frac{1}{2}\hat{\sigma}^{\alpha} \otimes \left(\begin{array}{cc}1 & 0 \\ 0 & 1\end{array}\right), \>\>\>\>\>\hat{s}_{2}^{\alpha} = \frac{1}{2}\left(\begin{array}{cc}1 & 0 \\ 0 & 1\end{array}\right) \otimes \hat{\sigma}^{\alpha},\\
\hat{\sigma}^{x} &= \left(\begin{array}{cc}0 & 1 \\ 1 & 0\end{array}\right), \>\>\>\> \hat{\sigma}^{y} = \left(\begin{array}{cc}0 & -i \\ i & 0\end{array}\right),\>\>\>\> \hat{\sigma}^{z} = \left(\begin{array}{cc}1 & 0 \\ 0 & -1\end{array}\right),
\end{aligned}
\end{equation}
which yields,
\begin{equation}
\hat{H}_{\mathrm{Spin}}^{(2)} = V_{\mathrm{Z}} \left(\hat{s}_{1}^{x} \hat{s}_{2}^{x} + \hat{s}_{1}^{y} \hat{s}_{2}^{y} + \Delta_{\mathrm{Z}}\hat{s}_{1}^{z}\hat{s}_{2}^{z}\right) + \Omega_{\mathrm{Z}}(\hat{s}_{1}^{z} + \hat{s}_{2}^{z}) + \Omega(\hat{s}_{1}^{z} + \hat{s}_{2}^{z}),
\end{equation}
with $V_{\mathrm{Z}}$, $\Delta_{\mathrm{Z}}$, $\Omega_{\mathrm{Z}}$ defined as in the main text.

We can extrapolate this model to a full lattice by replacing the subscripts $1$, $2$ with $\vec{r}$, $\vec{r}+\vec{\mathrm{Z}}$ and summing over all $\vec{r}$. The superexchange interactions along the other lattice directions $\mathrm{X}$, $\mathrm{Y}$ are computed analogously, except setting $\Omega = G = \phi = 0$. The resulting Hamiltonian is $\hat{H}_{\mathrm{Spin}}$ in the main text. Note that the last term $\Omega (\hat{s}_{1}^{z} + \hat{s}_{2}^{z})$ is the laser drive, which we write directly as $\Omega \sum_{\vec{r}} \hat{s}_{\vec{r}}^{z}$ (rather than replacing subscripts) to avoid double counting.

\subsection{Bosonic superexchange model}
\label{app_Bosonic}

While the main text studied mostly fermionic atoms, we can also derive a superexchange model for bosons. The underlying Hubbard Hamiltonian $\hat{H}$ describing the system is very similar. The annihilation operators $\hat{c}_{\vec{r},\sigma}$ and $\hat{a}_{\vec{r},\tilde{\sigma}}$ will now have bosonic commutation relations rather than fermionic, and the on-site interaction will instead be written as,
\begin{equation}
    \hat{H}_{\mathrm{Interaction}}= \frac{U}{2}\sum_{\vec{r}}\hat{n}_{\vec{r}}\left(\hat{n}_{\vec{r}}-\mathbbm{1}\right),
\end{equation}
where $\hat{n}_{\vec{r}}= \hat{n}_{\vec{r},g}+\hat{n}_{\vec{r},e}$. In writing this interaction we have assumed that the scattering lengths of collisions between $ee$, $eg$, $gg$ pairs are all the same (in contrast to the fermionic case which only has one characteristic singlet scattering length). Non-equal scattering lengths can still be accounted for with the Schrieffer-Wolff approach, leading to corrections in the spin model coefficients; we delegate such calculations to future work. Everything else about the models and mappings is the same.

Using the bosonic version of the Hubbard model, we can derive superexchange interactions in an analogous procedure. There will be more basis states for two neighbouring sites (namely states $\ket{2,0,0,0}$, $\ket{0,2,0,0}$, $\ket{0,0,2,0}$, $\ket{0,0,0,2}$) that need to be traced out, but otherwise the derivation proceeds the same way. We find the following superexchange interactions:
\begin{equation}
\begin{aligned}
\hat{H}_{\mathrm{Spin}}^{\mathrm{B}} = - \frac{4J_{\mathrm{X}}^2}{U}&\sum_{\vec{r}} \vec{s}_{\vec{r}}\cdot \vec{s}_{\vec{r}+\vec{\mathrm{X}}} - \frac{4J_{\mathrm{Y}}^2}{U}\sum_{\vec{r}}\vec{s}_{\vec{r}}\cdot \vec{s}_{\vec{r}+\vec{\mathrm{Y}}}\\
-V_{\mathrm{Z}}&\sum_{\vec{r}}\left[\vec{s}_{\vec{r}}\cdot \vec{s}_{\vec{r}+\vec{\mathrm{Z}}} + \left(\Delta_{\mathrm{Z}}-1\right)\hat{s}_{\vec{r}}^{z}\hat{s}_{\vec{r}+\vec{\mathrm{Z}}}^{z}\right]+\sum_{\vec{r}}\left(\Omega_{\mathrm{Z+}}\hat{s}_{\vec{r}}^{z} + \Omega_{\mathrm{Z-}}\hat{s}_{\vec{r}+\vec{\mathrm{Z}}}^{z}\right) + \Omega \sum_{\vec{r}}\hat{s}_{\vec{r}}^{z}.
\end{aligned}
\end{equation}
All of the interactions have the same magnitude but the opposite sign compared to the fermionic case. The explicit forms of the parameters $V_{\mathrm{Z}}$, $\Delta_{\mathrm{Z}}$ are the same as in the main text. The only other difference is the superexchange single-particle shifts, which have coefficients of:
\begin{equation}
    \Omega_{\mathrm{Z}\pm} = -\frac{2J_{\mathrm{Z}}^2 \Omega \left(U^2-\Omega^2 +G^2 \pm 4 U G\right)}{(U^2-\Omega^2-G^2)^2-4\Omega^2 G^2} \sin^2 \left(\frac{\phi}{2}\right).
\end{equation}
These single-particle shifts still only have non-trivial contribution at the system boundaries, so they will be negligible for large enough systems, and can also be mitigated with dynamical decoupling pulses as discussed in the main text.

\section{Realistic experimental parameters}
\label{app_Parameters}

In this Appendix we collect and clarify the various experimental parameters that are relevant to our proposed protocol for a 3D optical lattice implementation. The base lattice parameters are:
\begin{itemize}
\item The atomic mass $m$.
\item The lattice laser wavelength $\lambda_{\mathrm{L}}$, which sets the lattice spacing $a = \lambda_{\mathrm{L}}/2$.
\item The wavelength $\lambda_{\mathrm{d}}$ of the laser driving spin-flips between $e$ and $g$.
\end{itemize}
From these parameters we obtain:
\begin{itemize}
\item The gravity per site $G = m g a$, with $g$ the local gravitational acceleration.
\item The spin-orbit phase $\phi = 2\pi a/\lambda_{\mathrm{d}}$ if using a direct (single-photon) optical transition between $e$ and $g$, such as a clock transition. For a Raman (two-photon) transition between e.g. hyperfine states, $\phi = 4\pi a/\lambda_{\mathrm{d}}$ assuming the Raman beams are counter-propagating along $\mathrm{Z}$.
\end{itemize}
The next relevant set of parameters is the lattice depths along all directions,
\begin{itemize}
\item The lattice depths $(v_{\mathrm{X}},v_{\mathrm{Y}},v_{\mathrm{Z}})$ typically measured in units of the recoil energy $E_{\mathrm{R}}=\hbar^2 \pi^2 / (2m a^2)$.
\end{itemize}
These depths determine the lowest-band tunneling and interaction parameters,
\begin{itemize}
\item The tunneling rates,
\begin{equation}
J_{\nu} = \int_{-\infty}^{\infty}d\nu \>w_{0}^{*}(\nu)\left[-\frac{\hbar^2}{2m}\frac{d^2}{d\nu^2} + v_{\nu} E_{\mathrm{R}} \sin^2 \left(\frac{\pi \nu}{a}\right)\right]w_{0}(\nu-a),
\end{equation}
with $w_0(\nu)$ the lowest-band Wannier function for the lattice along $\nu$.
\item The on-site interactions,
\begin{equation}
    U = \frac{4\pi \hbar^2 a_{s}}{m} \int_{-\infty}^{\infty} d\mathrm{X} \int_{-\infty}^{\infty} d\mathrm{Y}\int_{-\infty}^{\infty} d\mathrm{Z} \> |w_0(\mathrm{X})|^4 |w_0(\mathrm{Y})|^4 |w_0(\mathrm{Z})|^4,
\end{equation}
with $a_{s}$ the $s$-wave scattering length between the $e$ and $g$ atoms.
\end{itemize}
From the above, we obtain the strength of superexchange interactions along all directions,
\begin{itemize}
    \item Superexchange rates along $\mathrm{X}$ and $\mathrm{Y}$,
\begin{equation}
    V_{\mathrm{X}} = \frac{4J_{\mathrm{X}}^2}{U},\>\>\>\>V_{\mathrm{Y}} = \frac{4J_{\mathrm{Y}}^2}{U}.
\end{equation}
    \item Superexchange rates along $\mathrm{Z}$,
\begin{equation}
    V_{\mathrm{Z}} = \frac{4J_{\mathrm{Z}}^2 U}{U^2-G^2}\cos^2 \left(\frac{\phi}{2}\right).
\end{equation}
\end{itemize}
The only remaining free parameter is the Rabi frequency $\Omega$ of the driving laser, which can be used to tune the anisotropy of the $\mathrm{Z}$ interactions:
\begin{equation}
\Delta_{\mathrm{Z}}=1- \frac{(U^2-G^2)(U^2-\Omega^2-G^2)}{(U^2-\Omega^2-G^2)^2-4\Omega^2G^2}\tan^2 \left(\frac{\phi}{2}\right).
\end{equation}

\begin{figure}[h]
\includegraphics[width=0.7\textwidth]{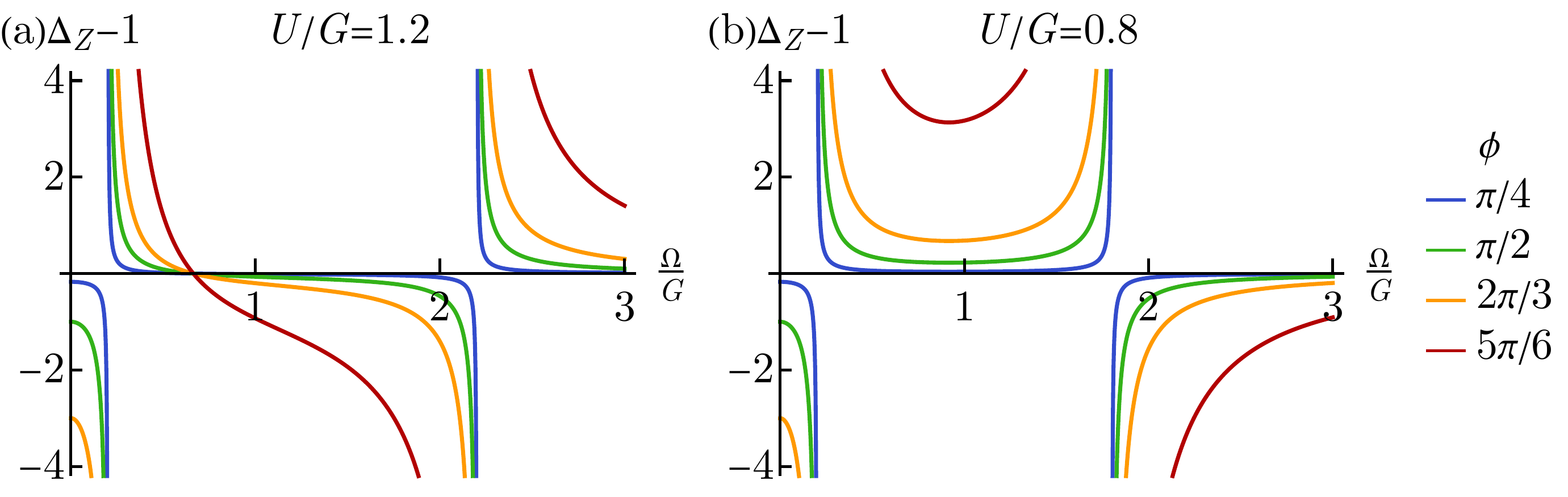}
\centering
\caption{Anisotropy for different values of flux $\phi$ as a function of drive Rabi frequency $\Omega$, for (a) interactions stronger than gravity $U > G$ and (b) weaker than gravity $U < G$.}
\label{fig_AnisotropyExperimental}
\end{figure}

In Fig.~\ref{fig_AnisotropyExperimental} we plot the anisotropy for two regimes $U > G$ and $U < G$ (which roughly correspond to deeper and shallower lattices respectively). The deeper lattice regime offers an easily tunable range of anisotropies for drive Rabi frequencies $\Omega \sim G$ that are not too strong for experimentally realistic scales. The shallower lattice regime does not have anisotropy that is as easily tunable for a fixed SOC phase $\phi$, but offers faster superexchange timescales.

We provide two sets of sample parameters for realistic candidate systems in Table~\ref{tab_Parameters}. The first candidate system is \textbf{(A)} fermionic $^{87}$Sr using $^{1}\mathrm{S}_0$ and $^{3}\mathrm{P}_0$ clock states as spin states $g$, $e$, in a magic-wavelength optical lattice. The second candidate is \textbf{(B)} bosonic $^{87}$Rb using two magnetic field-insensitive hyperfine states with total angular momentum projection $M_F = 0$ in the ground electronic manifold as spin states $e$, $g$. These candidate systems and parameters are not meant to be restrictive or exhaustive; they are provided just to have a sense of the energy scales involved.

\begin{table}[h]
\label{tab_Parameters}
\begin{tabular}{|c|c|c|c|c|c|c|c|c|c|c|c|}
\hline
System & $\lambda_{\mathrm{L}}$ [nm] & $\lambda_{\mathrm{d}}$ [nm] & $G$ [kHz] & $\phi$ & $(v_{\mathrm{X}}, v_{\mathrm{Y}}, v_{\mathrm{Z}})$ [$E_{\mathrm{R}}$] & $(J_{\mathrm{X}},J_{\mathrm{Y}},J_{\mathrm{Z}})$ [Hz] & $U$ [kHz] & $V_{\mathrm{XY}}$ [Hz] & $V_{\mathrm{Z}}$ [Hz] & $\Omega$ [kHz] & $\Delta_{\mathrm{Z}}-1$ \\ \hline
\textbf{(A)} & 813 & 698 & 0.87 & 1.17$\pi$ & (14,14,16) & (28,28,18) & 0.94 & 3.3 & 0.7 & 1.28 & -1 \\ \cline{11-12} 
& & &  &  &  &  &  &  &  & 0.22 & 1 \\ \hline
\textbf{(B)} & 1064 & 795 & 1.14 & 2.68$\pi$ & (14,14,12) & (16,16,25) & 0.55 & 1.9 & -0.3 & 0.3 & -5 \\ \cline{11-12}
& & & & & & & & & & 1.1 & 5 \\ \cline{11-12} 
& & & & & & & & & & 2.6 & -1 \\ \hline
\end{tabular}
\caption{Sample parameters for two sample systems. System \textbf{(A)} is fermionic $^{87}$Sr, using clock states $^{1}\mathrm{S}_0$, $^{3}\mathrm{P}_0$ with transition frequency $\lambda_{\mathrm{d}}=698$ nm as spin states in a $\lambda_{\mathrm{L}}=813$ nm magic-wavelength optical lattice. The scattering length is $a_{s} = 69a_0$. System \textbf{(B)} is bosonic $^{87}$Rb trapped in a conventional $\lambda_{\mathrm{L}}=1064$ nm lattice, using ground-electronic hyperfine states $\ket{\mathrm{F}=2, M_\mathrm{F} = 0}$, $\ket{\mathrm{F}=3, M_\mathrm{F} = 0}$ as spin states coupled by a Raman transition with wavelength $\lambda_{\mathrm{d}}=795$ nm. The scattering length is $a_s = 98 a_0$ (assumed to be equal for all three scattering channels $gg$, $eg$, $ee$).}
\end{table}

\end{document}